\documentclass{ieeetmlcn}
\usepackage{cite}
\usepackage{amsmath,amssymb,amsfonts, amsthm, bm, bbm}
\usepackage{algorithmic}
\usepackage{color}
\usepackage{graphicx}
\usepackage{textcomp}
\usepackage{xcolor}
\usepackage{hyperref}
\usepackage{listings}
\usepackage{algorithm,algorithmic}
\usepackage{threeparttable}
\usepackage{tabularx}
\usepackage{subfigure}
\usepackage{float}
\usepackage{color}
\usepackage{adjustbox}
\usepackage{multicol}
\usepackage{dblfloatfix}
\usepackage{cuted, nccmath}
\usepackage{lipsum}
\usepackage{tablefootnote}

\usepackage{mdframed}
\hypersetup{
    colorlinks=true,
    linkcolor=black,
    citecolor=black,
    urlcolor=red,
}

\usepackage{esint}

\long\def\comment#1{}

\newtheorem{definition}{Definition}
\newtheorem{thm}{Theorem}

\def\figref#1{Fig.~\ref{#1}}
\def\be{\begin{equation} }
\def\ee{\end{equation}}

\def\BibTeX{{\rm B\kern-.05em{\sc i\kern-.025em b}\kern-.08em
    T\kern-.1667em\lower.7ex\hbox{E}\kern-.125emX}}
\AtBeginDocument{\definecolor{tmlcncolor}{cmyk}{0.93,0.59,0.15,0.02}\definecolor{NavyBlue}{RGB}{0,86,125}}

\def\OJlogo{\vspace{-4pt}\includegraphics[height=18pt]{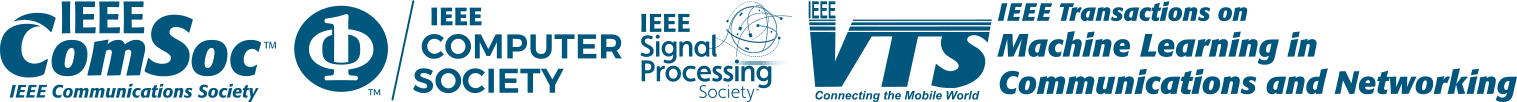}}
\def\seclogo{\vspace{10pt}\includegraphics[height=18pt]{new_logo.png}}

\def\authorrefmark#1{\ensuremath{^{\textbf{#1}}}}

\begin{document}
\receiveddate{XX Month, XXXX}
\reviseddate{XX Month, XXXX}
\accepteddate{XX Month, XXXX}
\publisheddate{XX Month, XXXX}
\currentdate{XX Month, XXXX}
\doiinfo{TMLCN.2022.1234567}

\markboth{}{Author {et al.}}

\title{Outage Performance and Novel Loss Function for an ML-Assisted Resource Allocation: An Exact Analytical Framework}

\author{Nidhi Simmons\authorrefmark{1}, Senior Member, IEEE, David E. Simmons\authorrefmark{2},\\ and Michel Daoud Yacoub\authorrefmark{3}, Member, IEEE}
\affil{Centre for Wireless Innovation, 
Institute of Electronics, Communications and Information Technology, Queen's University of Belfast, Belfast, BT3 9DT, UK, (email: nidhi.simmons@qub.ac.uk)}
\affil{Dhali Holdings Ltd., Belfast, BT5 7HW, UK, (email: dr.desimmons@gmail.com)}
\affil{Wireless Technology Laboratory,
School of Electrical and Computer Engineering, University of Campinas, Campinas 13083-970, Brazil, (email: mdyacoub@unicamp.br)}
\corresp{Corresponding author: Nidhi Simmons (email: nidhi.simmons@qub.ac.uk).}
\authornote{The first two authors have contributed equally to this work which is supported by the Royal Academy of Engineering (grant ref RF\textbackslash201920\textbackslash 19\textbackslash 191). 
}

\begin{abstract}
{In this paper, we present Machine Learning (ML) solutions to address the reliability challenges likely to be encountered in advanced wireless systems (5G, 6G, and indeed beyond). Specifically, we introduce a novel loss function to minimize the outage probability of an ML-based resource allocation system. A single-user multi-resource greedy allocation strategy constitutes our application scenario, for which an ML binary classification predictor assists in selecting a resource satisfying the established outage criterium. While other resource allocation policies may be suitable, they are not the focus of our study. Instead, our primary emphasis is on theoretically developing this loss function and leveraging it to train an ML model to address the outage probability challenge.} With no access to future channel state information, this predictor foresees each resource's likely future outage status. When the predictor encounters a resource it believes will be satisfactory, it allocates it to the user.
 The predictor aims to ensure that a user avoids resources likely to undergo an outage. Our main result establishes exact and asymptotic expressions for this system's outage probability. 
 {
 These expressions reveal that focusing solely on the optimization of the per-resource outage probability conditioned on the ML predictor recommending resource allocation
(a strategy that - at face value - looks to be the most appropriate) may produce inadequate predictors that reject every resource. They also reveal that focusing on standard metrics, like precision, false-positive rate, or recall, may not produce optimal predictors.
 }
 With our result, we formulate a theoretically optimal, differentiable loss function to train our predictor. We then compare predictors trained using this and traditional loss functions namely, binary cross-entropy (BCE), mean squared error (MSE), and mean absolute error (MAE). In all scenarios, predictors trained using our novel loss function provide superior outage probability performance. Moreover, in some cases, our loss function outperforms predictors trained with BCE, MAE, and MSE by multiple orders of magnitude. {Additionally, when applied to another ML-based resource allocation scheme (a modified greedy algorithm), our proposed loss function maintains its efficacy.}

\end{abstract}



\maketitle

\section{INTRODUCTION}

\IEEEPARstart{W}{ireless} channels are inherently dynamic, strongly influenced by their surrounding environments and users' mobility. Their ability to support communication fluctuates with space, time, and frequency. Such random dynamicity directly impacts the quality of service experienced by the user. Hence, sophisticated control techniques are pivotal to minimizing the deterioration posed by unfavorable conditions. 

Advanced generations of wireless systems, {such as 6G, and beyond} will predominantly operate at millimeter wave (mmWave) and terahertz (THz) bands as this is one of the proposed solutions to handle the ever-increasing data traffic demands caused by the massive growth of {connected devices~\cite{tripathi2021millimeter}. High radio frequencies make communication channels more susceptible to interference from surroundings, such as obstacles and atmospheric conditions, which would have minimal impact on links operating at lower frequencies.
 Hence signal degradation and unstable communications may become frequent~\cite{sarieddeen2020next, 8254900, 9681870}. Moreover, in advanced systems, devices are expected to play an active role in decision-making (without human involvement) tasks. Thus, it is paramount that their actions be highly reliable. Operation reliability will be further amplified for critical, immersive, and omnipresent communications.}

{Indeed, it may be possible to increase the communication reliability if the actions are taken based on adequately predicting the future status of the link. To address these challenges, there is a requirement for agile, high-dimensional modelling techniques with real-time adaptation~\cite{ali20206g}. Machine Learning (ML) is becoming a pivotal tool in all walks of applications, and advanced wireless communications systems (5G, 6G, and indeed beyond) are no exception. Regarding the reliability of radio links in complex systems such as these, for which intertwined phenomena, parameters, and metrics are at play in a constantly changing scenario, guaranteeing the proper quality of service is a challenge. In cases like this, ML techniques arise as a handy and powerful instrument.
ML is key, offering flexibility by letting data, which describes the system and its performance, drive decision-making. Through ML, we can adapt more effectively to the dynamic nature of future systems, enabling seamless communication and enhanced problem-solving.} 

{Demonstrating its capabilities, ML has been effective in predicting outages, identifying link blockages, and assessing link quality~\cite{wu2022deep, 9112745, 9795301, 8968748, MOON20221, 9562750, 9129369, 9512383}.}  For example, in~\cite{MOON20221}, a deep neural network was trained to map user positions and data traffic demand to their corresponding blockage status and optimal beam index. It was shown that this scheme could predict blockages in mmWave communications with 90\% accuracy. Furthermore,~\cite{9562750} implemented a framework for predicting blockages in mmWave and THz systems based on meta-learning that trained a recurrent
neural network using just a few data samples. Other methods, e.g., combining computer vision and deep learning tools, were also used to identify blockages in~\cite{9129369} and~\cite{9512383}. 

Further expanding its utility, ML has been used to optimise resource allocation in wireless systems, and improve channel estimation in multiple-input multiple-output (MIMO) or in orthogonal frequency division multiplexing (OFDM) systems~\cite{ye2019deep, sanguinetti2018deep, 9048929, 9237143, wen2018deep}. For instance,~\cite{ye2019deep} proposed a deep reinforcement learning based approach for resource allocation in vehicle-to-vehicle communications. The developed algorithm could be applied to unicast and broadcast scenarios, and reduced interference in vehicle-to-infrastructure communications. 
In~\cite{sanguinetti2018deep}, the potential of deep learning for power allocation in massive MIMO systems was investigated, demonstrating a significant reduction in complexity and processing time of the optimization process, while~\cite{9048929} presented a novel concept of employing deep neural networks to learn the channel-to-channel mapping between different sets of antennas and frequency bands, with the aim of optimizing performance in similar systems.
In \cite{9237143}, deep reinforcement learning was employed to optimize energy efficiency in device-to-device enabled heterogeneous networks.

{In the studies highlighted above, the underlying learning algorithms are often directed by traditional loss functions, such as binary cross-entropy (BCE) or mean squared error (MSE). At its core, a loss function evaluates the gap between the predicted outputs of the model and the actual observations. This evaluation guides the learning algorithm to minimize this difference by refining the model's parameters, commonly using approaches like gradient descent. However, our findings indicate that relying solely on these traditional loss functions can lead to marginal performance gains, potentially failing to satisfy the stringent demands of next-generation systems. Highlighting the necessity for a change in learning strategies,~\cite{9369424} emphasized the significance of incorporating domain-specific knowledge from communication systems into the learning procedure.}

{Recently, there has been a growing momentum towards this domain-centric integration, aiming for optimal system performance. To this end, tailored loss functions are often developed to ensure that the learning process meets certain constraints or highlights specific prediction aspects, refining the model to better meet the unique requirements of a specific task. As an illustration, \cite{9130130} introduced three deep neural networks to approximate singular value decomposition in MIMO systems. The authors presented a tailored loss function for hybrid beamforming, considering the challenges of finite-precision phase shifters and power limitations. In~\cite{10094043}, a deep neural network-driven resource allocation method was proposed for cell-free massive MIMO systems with hardware limitations. This deep neural network employed a tailored loss function that was designed to optimize sum rates while taking into account user power and front-haul capacity limitations. The potential of reconfigurable intelligent surfaces (RISs) to enhance signal-to-noise ratio and improve network coverage using ML was explored in \cite{10166830}. Here, an unsupervised deep neural network with a tailored loss function was used to optimize RIS reflection coefficients.} 

{A generative adversarial network was used to improve wireless channel predictions in~\cite{9252921} by employing a tailored loss function that aimed at preserving low-rank channel matrices. Furthermore, a strategy for resource allocation was suggested in~\cite{9534716} for advanced network slicing in beyond 5G settings, using statistical federated learning coupled with a tailored loss function. Additionally, both \cite{9013851} and \cite{8761721} demonstrated the benefits of deep unsupervised and reinforcement learning methods, in their respective studies, for optimizing resource allocation in ultra-reliable low-latency communication using a tailored loss function.}


{Though these investigations offer important perspectives, they do not integrate outage probability reliability metric directly into their loss functions, specifically aiming to reduce the chances of link failures. This gap is significant, given that most link failures arise at the extreme, less probable ends of the channel's distribution (also referred to as deep-tail learning\footnote{{Recent years have witnessed a growing interest in deep-tail learning across various ML domains~\cite{zhang2023deep}.}}~\cite{zhang2023deep}). Failing to effectively learn from these statistically infrequent events during model training can result in ML approaches that are inadequately designed to mitigate such failures—crucial for meeting the demanding reliability expectations of future wireless technologies.}

\subsection{CONTRIBUTIONS}

To reduce outages encountered in future wireless systems, in this paper we use ML to predict and avoid link deterioration. The scenario exercised is a single-user multi-resource greedy allocation system amenable to machine intelligence. Here, the resource allocation strategy uses an ML binary classification predictor, which anticipates the future outage status of each resource. The predictor traverses the available resources, stopping when it encounters a resource it believes to be satisfactory. 
The goal of the predictor is to prevent the allocation of an unsatisfactory resource, which would cause outage. The main challenge in the prediction task is that the predictor does not have access to any future channel state information. Instead, it must use its predictive capacity to infer future resource outages based on its historical state.

{It is worth highlighting that our primary aim is to theoretically develop and showcase a custom loss function designed to enhance the outage probability performance of ML-assisted wireless systems. To achieve this, we first analyze the resource allocation system described above and derive associated outage probability results. We then construct a theoretical framework (i.e., a custom loss function) to effectively train our predictor, targeting the minimization of outage probability, which involves deep-tail learning.}
The main contributions are summarized as follows:
\begin{enumerate}
    \item Novel expressions are derived for the outage probability of a single-user multi-resource greedy allocation system which uses an ML binary classification predictor for detecting outages.
    \item  {These expressions reveal that focusing solely on the optimization of the per-resource outage probability conditioned on the ML predictor recommending resource allocation may produce inadequate predictors that reject every resource. Also, focusing on standard metrics, like precision, false-positive rate, or recall, may again not produce optimal predictors.}
    \item Leveraging the exact and asymptotic outage expressions derived for this resource allocation system, a novel custom loss function is formulated.
    \item A simulations-based assessment of our novel loss function is performed, and several valuable insights are obtained. Crucially, it is shown that this novel loss function significantly dominates the conventional loss functions. In some situations, it outperforms predictors trained using the BCE, MSE and mean absolute error~(MAE)~\cite{hastie2009elements} by approximately two orders of magnitude (i.e., $100\times$).
    \item {It is also demonstrated that our custom loss function is particularly effective in regions experiencing infrequent outages, illustrating its capability to facilitate deep-tail learning.}
    \item {Finally, even when applied to a different ML-based resource allocation scheme (modified greedy algorithm), we show that our proposed loss function retains its effectiveness.}
\end{enumerate}

\subsection{ORGANIZATION AND NOTATION}
Section~\ref{sec:system_model} describes the system model, which consists of the resource allocation system combined with an ML binary classification predictor. Section~\ref{sec:primary_results} provides exact and asymptotic expressions for this system's outage behaviour. Section~\ref{sec:applications} builds on the theory presented in Section~\ref{sec:primary_results} to construct a novel loss function that is used to train an ML predictor for our system. Section~\ref{sec:sims} presents experimental results using our custom loss function. Finally, the work is concluded in Section~\ref{sec:conc}.
Table~\ref{tab:notation} highlights important notation used throughout the paper.
\begin{table*} 
\small
\centering
\renewcommand{\arraystretch}{1.03}
\caption{Notations}
\label{table1}
\begin{threeparttable}
\begin{tabular}{|c|c|c|}
\hline 
Notation & Description & Declaration\tabularnewline
\hline 
\hline 
$\mathcal{R}$ & The set of available resources &~Fig.~\ref{fig:system_model}\tabularnewline
\hline 
$h_{i}{(t)}$ & The $i$th resource's channel state & \eqref{eq:channel_state}\tabularnewline
\hline 
$H_{i}{(t, k)}$ & A $k$ length vector of $h_{i}{(t)}$ for period $t-k+1$ to $t$& \eqref{eq:H_Vector}\tabularnewline
\hline 
$C\left(H_{i}{(t, k)}\right)$ & Resource $i$'s channel capacity for period $t-k+1$ to $t$ & \eqref{eq:Gaussian_channel_capacity}\tabularnewline
\hline 
$\gamma_{th}$ & Rate threshold required by Alice & \eqref{eq:outage_probability_for_singe_resource}\tabularnewline
\hline
$P_{1}\left(\gamma_{th}\right)$ & Outage probability of the system with a single resource & \eqref{eq:outage_probability_for_singe_resource} \tabularnewline
\hline 
$Q\left(H_{i}{(t, k)};\Theta\right)$ & The ML binary classifier with output in $ [0,1]$; $\Theta$ denotes the predictor's parameters & \eqref{eq:model_definition}\tabularnewline
\hline 
$q_{th}$ & The threshold above which outages are predicted by the predictor & \eqref{eq:Qandqth}\tabularnewline
\hline 
$F_{Q}\left(x\right)$ & The cumulative distribution function of the predictor's output & \eqref{eq:Q_CDF}\tabularnewline
\hline 
$\mathcal{R}^\star(q_{th}) $ & The subset of resources for which the predictor predicts no outages &~\eqref{eq:Rstar}\tabularnewline
\hline 
$P_{\infty}\left(\gamma_{th}, q_{th}\right)$ & Outage probability of the system with $\left|\mathcal{R}\right|\to\infty$
resources & \eqref{eq:limiting_outage_def} \tabularnewline
\hline 
$P_{|\mathcal{R}|}\left(\gamma_{th}, q_{th}\right)$ & Outage probability of the system with $\left|\mathcal{R}\right|$
resources & \eqref{eq:main_outage_expression} or~\eqref{eq:main_outage_expression_2}
 \tabularnewline
\hline 
\end{tabular}
\end{threeparttable}
\label{tab:notation}
\end{table*}

\section{SYSTEM MODEL \label{sec:system_model}}
We present our system model in three parts. The first describes the channel model adopted for a resource. The second describes the ML binary classification predictor. The third describes the role of the predictor when allocating a resource to a user.

\subsection{CHANNEL MODEL FOR A RESOURCE\label{sec:channel_model}}

We consider a generic single-user multi-resource system. 
The resources could be, e.g., non-overlapping portions of the spectrum~\cite{wong2009optimal, vcabric2005cognitive, kwon2009generalized}, eigen-channels of a spatial multiplexing transceiver \cite{8491076},  relays in a multi-relay system~\cite{6601774}, etc. 

Resource $i\in\mathcal{R}$ is assumed to have a fluctuating channel state, which can be expressed as a time-series, 
\begin{equation}
    h_i(t)\in\mathbb{C}.\label{eq:channel_state}
\end{equation}
{
Each unit increment in $t$ corresponds to a channel sample interval. The sampling interval is a function of the correlation bandwidth (or time or distance), which in turn is a function of the operation frequency and the velocity of the receiver. Typically for a system operating at 5.8~GHz, the sampling interval can range from 0.1~ms to 1~ms~\cite{9795301, 7467556}.  
}

The state $h_i(t)$ is accessible by a user, Alice, before she is allocated the resource. 
It is assumed that $h_i(t)$ and $h_j(t)$ are independent and identically distributed\footnote{The i.i.d. assumption simplifies our theoretical analysis. Non-i.i.d. scenarios may be considered in the future.} (i.i.d.) for $i\neq j$. It is further assumed that $h_i(t)$ and $h_i(t + \Delta)$, $\Delta\in\mathbb{R}$, are identically distributed but may be correlated. 
This correlation declines as $\Delta$ increases and goes to zero as $\Delta\to\infty$. For our theoretical analysis, we do not impose any requirements on how the correlations decline.  Note that Alice has an ML classification predictor, the job of which will be to learn these correlations and aid her in selecting an appropriate resource for her future communication. We discuss this ML predictor and resource allocation strategy in Section~II.B and Section~II.C, respectively.

With $h_i(t)$ given by \eqref{eq:channel_state}, for $k\in\mathbb{N}$ we define the following vector of channel samples
\begin{equation}
{H_{i}}{(t , k)} \triangleq \left[h_{i}{(t-k+1)},\cdots, h_{i}{(t)} \right]^{T} \in \mathbb{C}^{k}.\label{eq:H_Vector}
\end{equation}
Here, $k$ is the length of the window of past samples of channel states for each resource $i \in \mathcal{R}$.
The ability for resource $i$ to support Alice's communication over a period $t-k+1$ to $t$ is determined by its capacity
$C\left({H_{i}}{(t , k)}\right) \in \mathbb{R}^+$
for that period~\cite{10.5555/1146355}. As an example, for a quasi-static Gaussian channel, {where $\mathtt{SNR}$ represents the average signal-to-noise ratio (SNR) per sample,} the capacity for resource $i$ in this period is given by~\cite[eq. (5.80)]{10.5555/1111206}
{
\begin{equation}
C\!\left({H_{i}}{(t , k)}\right)\! = \!\! \sum_{j=1}^{k}\log_2\!\left(\! 1 + \mathtt{SNR} \left| h_i\left(t\!-k\! + j\right)\right|^2\!\right) ~\rm{bits/s/Hz}.\label{eq:Gaussian_channel_capacity}
\end{equation}}
If Alice requires communication at a rate less than or equal to this capacity, the resource will be satisfactory for her. Otherwise, the resource will be in outage. For the remainder of this work, we assume that Alice's required communication rate is determined by a threshold $\gamma_{th}$.
The outage probability of a single resource $i$ can then be written as
{
\begin{equation}
    P_1\left( \gamma_{th} \right) \triangleq  \mathbb{P}  \left[C\left({H_{i}}{(t , k)}\right)< \gamma_{th} \right]~\forall ~ i\in\mathcal{R},
    \label{eq:outage_probability_for_singe_resource}
\end{equation}
where the equality across all $i$ is a consequence of the resources being i.i.d.}
\vspace{-0.5cm}
{\subsubsection*{Problem Statement:}
As mentioned previously,} Alice is equipped with an ML classification predictor whose goal is to allocate a resource $r^{\star} \in \mathcal{R}$ to her whilst trying to avoid outages. {In more detail, this ML predictor evaluates the current channel conditions $H_{i}(t,k)$ for a particular resource and predicts the likelihood of successful communication over the upcoming $l\in\mathbb{N}$ channel samples, $H_{i}(t+l,l)$, for that resource.} The main challenge here is that future channel state $h_{i}(t')$ for $t' > t$ is not available for any of the resources $i \in \mathcal{R}$. {Our resource allocation strategy hinges on the predictions made by this ML predictor to selectively allocate resources, ensuring minimal outages. Central to our study is to derive this system's outage probability and then establish a theoretical framework for constructing a communication system's loss function. This will optimize our predictor's training to reduce the probability of system outages. We now discuss the ML-based resource allocation predictor and the resource allocation strategy adopted here.}

\subsection{ML PREDICTOR FOR RESOURCE ALLOCATION}

Without loss of generality, it is assumed that resource $i$ begins at $i=1$, and for each time the predictor predicts an outage event, $i$ is incremented by $1$. 
For resource $i$, the predictor accepts an input vector of channel coefficients $H_{i}{(t,k)}$ defined in~\eqref{eq:H_Vector}. It provides an output value in the closed interval $[0,1]$, which is used to classify whether the following $l$ channel samples (i.e., 
{
$H_{i}{(t+l,l)}$
}) support communication without outage for each resource $i \in \mathcal{R}$.

{
This paper considers the \emph{general case} where the model may or may not be well calibrated. The \emph{special case} corresponding to well calibrated models should also satisfy the criteria~\cite{guo2017calibration}
\begin{equation}
    \mathbb{P}\left[ C\left({H_{i}}{(t+l , l)}\right) <  \gamma_{th} ~\vert~Q\left(H_{i}{(t,k)};\Theta\right) = q\right] \approx q~\forall ~q,\label{eq:calibration}
\end{equation}
where $\Theta$ represents the predictor's parameters and $q\in\left[0,1\right]$. The condition in~\eqref{eq:calibration} may or may not be the case for our analysis\footnote{This investigation may form part of our future work.}. This clarification implies that while the ML predictors in our study are designed to learn conditional probability distributions, they are not explicitly required to fulfill the \emph{additional} calibration condition indicated in \eqref{eq:calibration}.
}

\begin{figure}[t!]
\centering
\includegraphics[height=4.6cm, trim={10cm 7cm 11cm 1.5cm},clip]{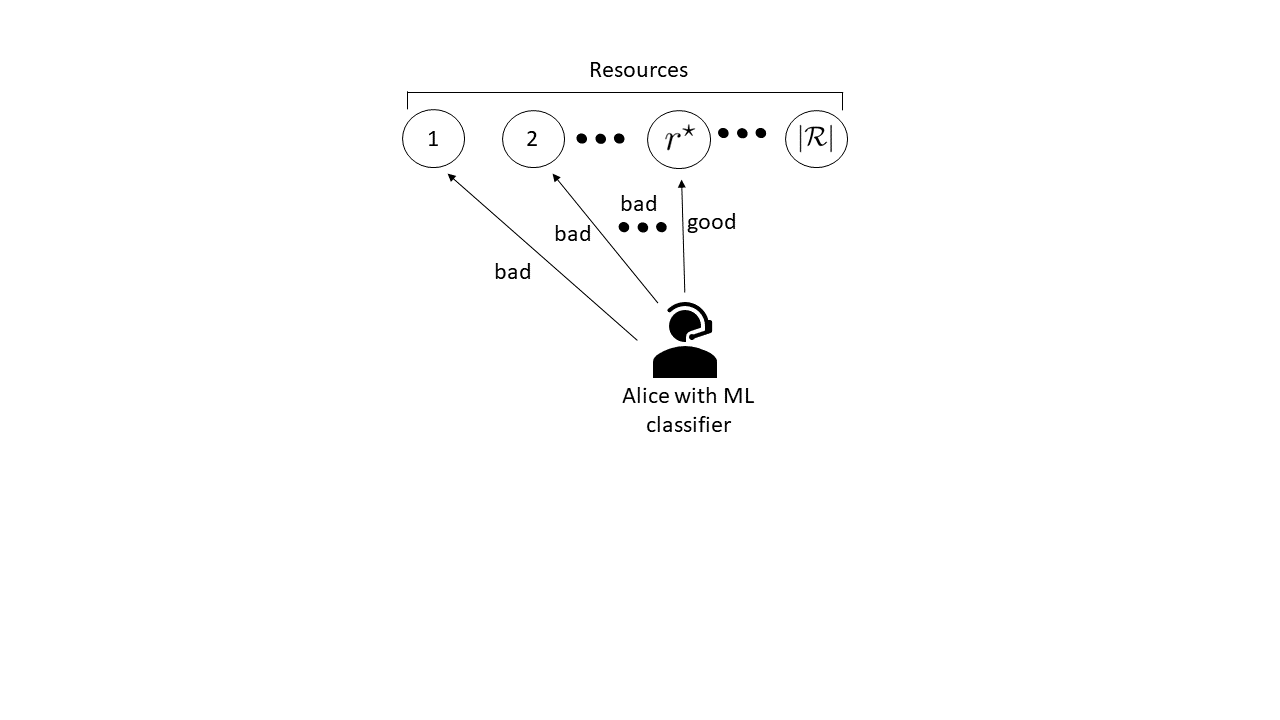}
\caption{An example of Alice accessing available resources. Here, she scans from left to right and chooses the first good resource, $r^{\star}$. These resources could be portions of the spectrum in a licensed or an unlicensed band, sub-bands in frequency-division multiple access (FDMA), subcarriers in an Orthogonal-FDMA (OFDMA) system, or eigen channels of a MIMO system, etc.}
\label{fig:system_model}
\end{figure}

In this work,
\begin{equation}
    Q\left(H_{i}{(t,k)};\Theta\right) \in \left[0,1\right],\label{eq:model_definition}
\end{equation}
For some choice of the predictor's classification threshold $q_{th}\in[0,1]$,
\begin{equation}
    Q\left(H_{i}{(t,k)};\Theta\right) > q_{th}\label{eq:Qandqth}
\end{equation}
implies that an outage is predicted for the next $l$ samples. Furthermore, the probability density function (PDF) for the predictor's output is given by
$  f_Q\left(x\right),
$
where 

\begin{multline}
    F_Q\left(x\right) \triangleq  \int_{x'=0}^{x}f_Q\left(x'\right)dx'\\ = \mathbb{P}\left[ Q\left(H_{i}{(t,k)};\Theta\right)\leq x  \right]~\forall~i\in\mathcal{R}\label{eq:Q_CDF}
\end{multline}
is the cumulative distribution function of the predictor's output,
{
and the equality across all $i$ is a consequence of the resources being i.i.d.} 
Importantly, for $x = q_{th}$, we retrieve the predictor's resource acceptance probability, $F_Q\left(q_{th}\right)$. Alternatively, the predictor's resource rejection probability is given by $1-F_Q\left(q_{th}\right)$. We next discuss the resource allocation strategy adopted here.

\subsection{RESOURCE ALLOCATION}
{To illustrate the feasibility of our novel loss function, we examine a basic communication setup and resource allocation method. Specifically, we employ a greedy resource allocation policy similar to those found in~\cite{nwamadi2008dynamic, 7036039, 4161913} as they are known for their efficient implementation and reduced time complexity. Our objective is not to explore different channel allocation techniques, although this can be done. Instead, our primary focus is on introducing a novel approach to construct loss functions for communication systems and evaluating their performance. Our preliminary findings indicate that our novel approach can significantly enhance wireless system reliability when compared to commonly used loss functions. Thus, future research will explore the application of this technique in more intricate resource allocation systems.}


Fig.~\ref{fig:system_model} shows the allocation policy we consider. In this, Alice scans the available resources $\mathcal{R}$ and selects a resource $r^{\star}$ based on whether an ML binary classification predictor $Q$ 
predicts that future communication will succeed on that resource. 
The search for an appropriate resource for Alice stops when the predictor predicts a no-outage event for that resource. 
When none of the resources are predicted to be satisfactory {(referred to as the critical scenario), we explore two alternatives: 
$\mathtt{Case~1.}$ Alice is allocated the final resource~$|\mathcal{R}|$, or
$\mathtt{Case~2.}$~Alice is allocated the resource with the lowest ML binary classification value (the best predicted resource seen).} 

Accordingly, for ${\mathcal{R}} = \{1,2,\cdots, \vert \mathcal{R} \vert \}$, where resource $i \in \mathcal{R}$ is the $i$th one being explored by the allocation procedure, we can summarize the adopted strategy as follows. If
\begin{equation}
    \mathcal{R}^\star(q_{th}) = \left\{i\in\mathcal{R}~\mathrm{s.t.}~ Q({H}_{i}{(t,k); \Theta}) \leq q_{th} \right\},\label{eq:Rstar}
\end{equation}
is the subset of resources for which the predictor predicts no outages, then the scheme chooses:
{
\begin{align}
\mathtt{Case}&\mathtt{~1\!:}~r^{\star}\!\left(q_{th}\right)=\begin{cases} 
      \underset{i\in\mathcal{R}^\star(q_{th})}{\min} ~ i & \mathcal{R}^\star(q_{th})\neq\emptyset \\
      |\mathcal{R}| & \mathrm{~~otherwise.}
   \end{cases}
\label{eq:rgreedy}\\
\mathtt{Case}&\mathtt{~2\!:}~r^{\star}\!\left(q{th}\right) =\!\! \begin{cases}
\underset{i\in\mathcal{R}^\star\!(q_{th})}{\min} ~ i & \mathcal{R}^\star(q_{th})\!\neq\emptyset \\
\underset{i\in\mathcal{R}}{\mathrm{argmin}} ~ Q(H_i(t,k)) & \mathrm{otherwise.}
\end{cases}
\label{eq:rgreedy_case_i}
\end{align}
}

In what follows, the outage probability for a system with $\vert \mathcal{R} \vert$ resources  is denoted as $P_{\vert \mathcal{R} \vert}\left(\gamma_{th}, q_{th}\right)$, and the  outage probability in the infinite resource limit, i.e.,  ${\vert \mathcal{R} \vert} \rightarrow \infty$, is given by
\begin{equation}
P_\infty \left(\gamma_{th}, q_{th}\right) \triangleq \lim_{\left| \mathcal{R} \right| \to \infty}P_{\vert \mathcal{R} \vert}  \left(\gamma_{th}, q_{th}\right).\label{eq:limiting_outage_def}
\end{equation}

For brevity, going forward, we may drop the predictor's dependency on $\Theta$ in our notation where appropriate. 
\section{THE SYSTEM'S OUTAGE PROBABILITY\label{sec:primary_results}}

In this section, we present the main result of this work, describing how the ML binary classification predictor affects Alice's outage probability. 


\subsection{GENERAL OUTAGE EXPRESSIONS\label{sec:main_Results}}

We begin with the following theorem. 

\begin{thm}
Consider the resource allocation system described in Section~\ref{sec:system_model}, where $P_{1}\left(\gamma_{th}\right)$, $F_{Q}\left({q_{th}}\right)$ and $P_{\infty}\left(\gamma_{th}, q_{th}\right)$ are defined in~\eqref{eq:outage_probability_for_singe_resource}, \eqref{eq:Q_CDF}, and \eqref{eq:limiting_outage_def}, respectively. Furthermore, assume that all resources $i \in \mathcal{R}$ are i.i.d. Then, the outage probability for a system with $\vert \mathcal{R} \vert$ resources can be expressed as 

{
\begin{align}
\mathtt{Case }&~\mathtt{1}: P_{\vert \mathcal{R} \vert} \left(\gamma_{th}, q_{th}\right) \nonumber\\
=&~P_1 \left(\gamma_{th}\right) \left(1-F_Q\left( q_{th}\right)\right)^{\left| \mathcal{R} \right| - 1} \nonumber \\ 
&+ P_\infty \left(\gamma_{th}, q_{th}\right)\left(1 - \left(1 -  F_Q\left( q_{th}\right)\right)^{\left| \mathcal{R} \right| - 1} \right) 
 ,\label{eq:main_outage_expression}\\
\mathtt{Case }&~\mathtt{2}: P_{\vert \mathcal{R} \vert} \left(\gamma_{th}, q_{th}\right) \nonumber\\
=&~\mathbb{P}\left[ C(H_{i}{(t+l,l)})\!<\gamma_{th}~|~i = \underset{j\in\mathcal{R}}{\mathrm{argmin}} ~ Q(H_j(t,k)) \right]  \nonumber \\ 
&\times \left(1-F_Q\left( q_{th}\right)\right)^{\left| \mathcal{R} \right| - 1}\nonumber\\
&+ P_\infty \left(\gamma_{th}, q_{th}\right)\left(1 - \left(1 -  F_Q\left( q_{th}\right)\right)^{\left| \mathcal{R} \right| - 1} \right) ~\forall ~ i\in\mathcal{R}
 ,\label{eq:main_outage_expression_2}
\end{align}
}
where 
{
\begin{multline}
P_\infty \!\left( \gamma_{th}, q_{th}\right)\! \\= \mathbb{P}\left[ C(H_{i}{(t+l,l)})\!<\gamma_{th} |Q\!\left(H_{i}{(t,k)} \right) \!\leq q_{th} \right]~\forall ~ i\in\mathcal{R}.  \label{eq:infinite_outage_prob}
\end{multline}
}
{For \eqref{eq:main_outage_expression_2} and \eqref{eq:infinite_outage_prob}, the equality across all $i$ is a consequence of the i.i.d. resources.}

\label{thm thm1}
\end{thm}
\begin{proof}
See Appendix~\ref{app:A0}.
\end{proof}

{
A striking observation from this result, and one that we will discuss later, is that if we wish to minimise the outage probability of the system, we should not necessarily focus on minimising the outage probability of a resource given the predictor recommends allocating that resource, $P_\infty \!\left( \gamma_{th}, q_{th}\right)$. Rather, it is important that we also consider the number of resources available, the probabilities associated with rejecting them all, and the unconditional probability that a single resource is in outage. As we shall see later (discussions following Theorem~\ref{thm thm2}), if we fail to make such consideration, focusing only on minimising  the outage probability of a resource given the predictor recommends allocating that resource, we may end up in the unfortunate situation where our predictor refuses to select any resource at all.
}

From~\eqref{eq:main_outage_expression}, the outage probability for {$\mathtt{Case~1}$} can be viewed as the average over the limiting instantiations of the system when $\vert \mathcal{R} \vert$ $= 1$ and $\vert \mathcal{R} \vert$ $= \infty$, where the probabilities of the instantiations are, respectively, $ \left(1 - F_Q\left( q_{th}\right)\right)^{\left| \mathcal{R} \right| - 1}$ (i.e., the probability that all resources would be rejected by the predictor) and $1 - \left(1 -  F_Q\left( q_{th}\right)\right)^{\left| \mathcal{R} \right| - 1}$ (i.e., the probability that at least one resource would be accepted by the predictor). 
{From~\eqref{eq:main_outage_expression_2}, the outage probability {for {$\mathtt{Case~2}$}} can similarly be viewed as the average over the instantiations of the system when the index with the lowest predictor score is selected and when the system has $\vert \mathcal{R} \vert$ $= \infty$ resources.}

{For both cases} it is clear that when $\vert \mathcal{R} \vert$ $= 1$ or $\infty$, the system's outage probability  reduces to $P_1 \left(\gamma_{th}\right)$ and $P_{\infty} \left(\gamma_{th}, q_{th}\right)$, respectively, as expected.
{It can also be seen that the outage probability for the two cases differ only in their first term. Consequently, as this first term becomes small (relative to the second term) we expect the two cases to provide similar performance. Notably, this first term decays exponentially with $|\mathcal{R}|$. 
}

Furthermore, {
for $\mathtt{Case~1}$,
} inspecting the limiting values of $q_{th}=0$ and $1$,
we observe that $$P_{\vert \mathcal{R} \vert} \left(\gamma_{th}, q_{th}\right) \vert_{q_{th}=0,1} = P_1 \left(\gamma_{th}\right).$$ 
To see this, note that  $F_Q(q_{th})=0,1$ for $q_{th}=0,1,$ respectively. Thus,~\eqref{eq:main_outage_expression} reduces to $P_1 \left(\gamma_{th}\right)$ when $q_{th}=0$ and  $P_{\infty} \left(\gamma_{th}, q_{th}\right)$ when $q_{th} = 1$. Moreover, because $Q({H}_{i}{(t,k); \Theta}) \leq q_{th}$ is always true when $q_{th} = 1$, $P_{\infty} \left(\gamma_{th}, q_{th}\right)$ reduces to $P_1 \left(\gamma_{th}\right)$.
Again, considering \eqref{eq:Qandqth}, this behaviour should be expected because $q_{th}=0$ or $1$ will, respectively, result in the user always being allocated to the final or first resource {when $\mathtt{Case~1}$ is considered.}
{
For $\mathtt{Case~2}$, however, inspecting the limiting values of $q_{th}=0$ and $1$,
we observe different values for the outage probability. Specifically, we have \begin{align}
    P_{\vert \mathcal{R} \vert} & \left(\gamma_{th}, q_{th}\right) \vert_{q_{th}=0} \nonumber\\
    =& \mathbb{P}\left[ C(H_{i}{(t+l,l)})\!<\gamma_{th}~|~i = \underset{j\in\mathcal{R}}{\mathrm{argmin}} ~ Q(H_j(t,k)) \right]\label{eq:best_of_n}
\end{align}
and
\begin{align}
P_{\vert \mathcal{R} \vert} \left(\gamma_{th}, q_{th}\right) \vert_{q_{th}=1} = P_1 \left(\gamma_{th}\right).
\end{align}
From \eqref{eq:best_of_n} we see that $\mathtt{Case~2}$ reduces to the scenario in which the best predicted resource is always selected from the available resources $\mathcal{R}$.
}

{
Finally, it is noteworthy that the predictor's quality (i.e., the probability of communication success given predicted success) is exactly the complement of the system's outage probability in the infinite resource limit. This is because, with an infinite number of resources, a resource will almost surely be predicted to be good. And when it is, the outage probability observed by the system will simply be the probability that this selected resource is in outage. Because the resource is selected only when $Q\!\left(H_{i}{(t,k)} \right) \!\leq q_{th}$, we obtain \eqref{eq:infinite_outage_prob}. This is not true when only finitely many resources are available. Consequently, the outage probability in the infinite resource scenario reduces to the conditional expression in \eqref{eq:infinite_outage_prob}.
}

\subsection{OUTAGE EXPRESSIONS IN TERMS OF THE PREDICTOR'S OUTPUTS\label{sec:alternative_expressions}}

We next show that $P_{1}\left(\gamma_{th}\right)$,  $F_Q\left( q_{th} \right)$ and {$P_{\infty}\left(\gamma_{th}, q_{th}\right)$}, given in~\eqref{eq:outage_probability_for_singe_resource},~\eqref{eq:Q_CDF} and \eqref{eq:infinite_outage_prob}, respectively, can be expressed in terms of the predictor's true positives (TP), false positives (FP), true negatives (TN), and false negatives (FN). Critically, these alternative expressions will be used to formulate a differentiable loss function for training $Q$, which will be shown to be extremely effective at optimizing the system's performance. We begin with the following definition and key remark, which helps formalise TN, TP, FN, and FP.

\begin{definition}
Consider resource $i\in\mathcal{R}$. Also, let $f: \mathbb{R}\to [0,1]$ and $\mathcal{W}_n = \left\{ \left( H_{i}(t,k), b_{i}\right) \right\}$, $b_{i}\in \{0,1\}$, be a collection of $n$ labelled samples drawn uniformly and independently from our resource model described in Section~\ref{sec:system_model}, where label $b_{i}=0$ or $1$ implies, respectively, that $C\left(H_{i}(t+l,l)\right) \geq \gamma_{th}$ or $C\left(H_{i}(t+l,l)\right) < \gamma_{th}$ (i.e., a no-outage event or outage event follows $H_{i}(t,k)$). Then we define the following functionals:

\begin{align}
\mathrm{TN}\left( \mathcal{W}_n; f\right) \triangleq & \sum_{\mathcal{W}_n} f\left( q_{th} - Q\left(H_{i}{(t,k)}\right)\right)\left(1 - b_{i}\right)\label{eq:TN}\\
\mathrm{FN}\left( \mathcal{W}_n; f\right) \triangleq & \sum_{\mathcal{W}_n} f\left( q_{th}- Q\left(H_{i}{(t,k)}\right)\right)b_{i}\\
\mathrm{TP}\left( \mathcal{W}_n; f\right) \triangleq & \sum_{\mathcal{W}_n}  f\left(Q\left(H_{i}{(t,k)}\right) - q_{th} \right) b_{i} \label{eq:TP}\\
\mathrm{FP}\left( \mathcal{W}_n; f\right) \triangleq & \sum_{\mathcal{W}_n}  f\left(Q\left(H_{i}{(t,k)}\right) - q_{th} \right)\left(1 - b_{i}\right).\label{eq:FP}
\end{align}
\label{def:TPTNFPFN}
\end{definition}

\begin{figure}[t]
\centering
\includegraphics[width=5cm]{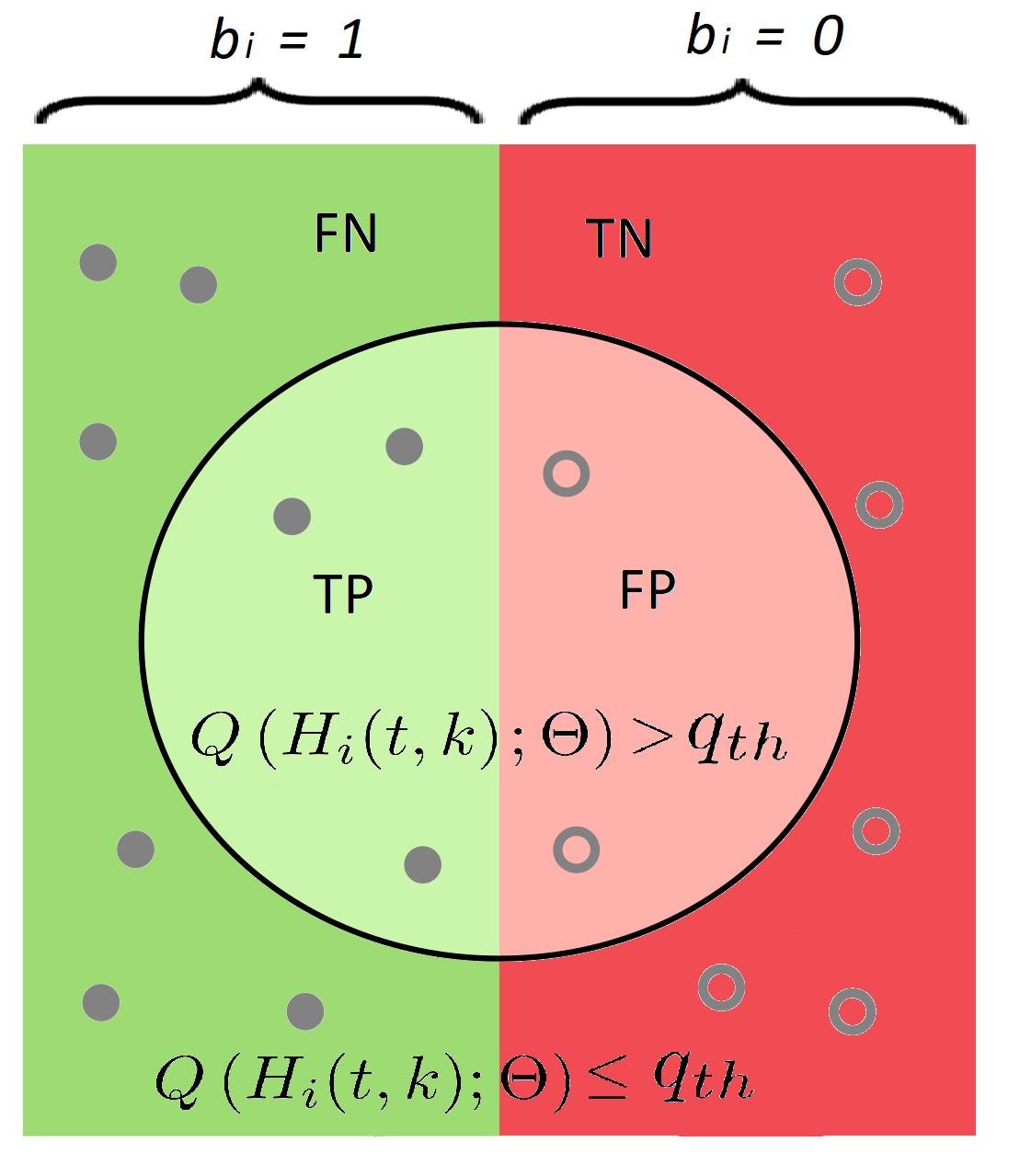}
\caption{Pictorial representation of samples from $\mathcal{W}_{n}$ (Definition \ref{def:TPTNFPFN}) along with an overlay identifying how each sample is classified by the predictor. Outages occur when $b_{i} = 1$, and predicted outages occur for all samples lying inside the circlular region. With the overlayed circular region, true positives (TP), true negatives (TN), false positives (FP) and false negatives (FN) can be identified.}
\label{fig:TPTNFPFN}
\end{figure}
Importantly, when $f(\cdot) = \mathbbm{1}(\cdot)$ (i.e., the Heaviside step function~\cite{referencewolfram_2022_unitstep}) in Definition~\ref{def:TPTNFPFN},~\eqref{eq:TN} - \eqref{eq:FP} reduce to an intuitive understanding of TN, FN, TP, and FP (see~Fig.~\ref{fig:TPTNFPFN}). In particular, the summands become a counter for TN, FN, TP and FP. For example, considering \eqref{eq:TN}, the Heaviside step function term is equal to $1$ \textit{only} when $Q(H_{i}{(t,k)}) \leq q_{th}$. When $Q(H_{i}{(t,k)}) \leq q_{th}$, the predictor is predicting no-outage. Moving to the second term $(1-b_{i})$ of \eqref{eq:TN}, this is equal to $1$ \textit{only} when the label $b_{i} = 0$. When $b_{i} = 0$, no-outage has occurred. Thus \eqref{eq:TN} is counting all the events when $Q(H_{i}{(t,k)}) \leq q_{th}$ and $b_{i} = 0$, i.e., when the predictor has predicted no-outage occurring when no-outages have occurred. These are the TNs. Likewise, considering \eqref{eq:TP}, the Heaviside step function term is equal to $1$ \textit{only} when $Q(H_{i}{(t,k)}) > q_{th}$. When $Q(H_{i}{(t,k)}) > q_{th}$, the predictor is predicting an outage. Moving to the second term $b_{i}$ of \eqref{eq:TP}, this is equal to $1$ when an outage has occurred. Thus \eqref{eq:TP} is counting all the events when $Q(H_{i}{(t,k)}) > q_{th}$ and $b_{i} = 1$, i.e., when the predictor has predicted an outage occurring when outages have occurred. These are the TPs.

Similar arguments apply to the other equations in Definition~\ref{def:TPTNFPFN} when $f(\cdot) = \mathbbm{1}(\cdot)$. The construction of \eqref{eq:TN} - \eqref{eq:FP} in this way will be helpful when formulating our custom loss function. Specifically, making a choice for $f(\cdot)$ that approximates $\mathbbm{1}(\cdot)$ and satisfies differentiable properties will allow us to effectively train a neural network predictor using back-propagation.

Alternative expressions are now presented in Theorem \ref{thm thm2} below for $P_{1}\left(\gamma_{th}\right)$, $F_Q\left( q_{th} \right)$, and $P_{\infty}\left(\gamma_{th}, q_{th}\right)$ given in~\eqref{eq:outage_probability_for_singe_resource}, \eqref{eq:Q_CDF}, and~\eqref{eq:infinite_outage_prob}, respectively. This allows us to express the system's outage probability {in $\mathtt{Case~1}$ purely} in terms of TN, TP, FN and FP. {However, the following theorem does not offer the same benefit for $\mathtt{Case~2}$, due to the complexities in formulating analogous equations. This constitutes an open problem for further investigation.}


\begin{figure*}[!ht]
 \begin{equation}
 \widehat{P}_{1}\left( \mathcal{W}_n;f\right) \triangleq \frac{\mathrm{TP}\left( \mathcal{W}_n;f\right) + \mathrm{FN}\left( \mathcal{W}_n;f \right)}{\mathrm{TN}\left( \mathcal{W}_n;f\right) + \mathrm{FN}\left( \mathcal{W}_n;f\right) + \mathrm{TP}\left( \mathcal{W}_n;f\right) + \mathrm{FP}\left( \mathcal{W}_n;f\right)},\label{eq:p_1_approx}
\end{equation}
\end{figure*}
\begin{figure*}[!ht]
\begin{equation}
    \widehat{ 
  F}_Q\left( \mathcal{W}_n;f\right)  \triangleq \frac{\mathrm{TN}\left( \mathcal{W}_n;f\right)+\mathrm{FN}\left( \mathcal{W}_n;f \right)}{\mathrm{TN}\left( \mathcal{W}_n;f\right)+\mathrm{FN}\left( \mathcal{W}_n;f\right) + \mathrm{TP}\left( \mathcal{W}_n;f\right)+\mathrm{FP}\left( \mathcal{W}_n;f\right)},\label{eq:q_approx}
\end{equation}
\end{figure*}
\begin{figure*}[!ht]
\begin{equation}
    \widehat{P}_{\infty}\left( \mathcal{W}_n;f\right) \triangleq \frac{\mathrm{FN}\left( \mathcal{W}_n;f\right)}{\mathrm{TN}\left( \mathcal{W}_n;f\right) + \mathrm{FN}\left( \mathcal{W}_n;f\right)}.\label{eq:r_approx}
\end{equation}
\hrulefill
\end{figure*}

\begin{thm}
Let $\mathcal{W}_n$, $\mathrm{TN}$, $\mathrm{TP}$, $\mathrm{FN}$ and $\mathrm{FP}$ be given by Definition~\ref{def:TPTNFPFN}. Furthermore, consider $\widehat{P}_{1}\left(\mathcal{W}_n;\mathbbm{1}\right)$, $\widehat{ 
  F}_Q\left(\mathcal{W}_n;\mathbbm{1} \right)$ and $ \widehat{P}_{\infty}\left(\mathcal{W}_n;\mathbbm{1}\right)$  given by \eqref{eq:p_1_approx}, \eqref{eq:q_approx} and \eqref{eq:r_approx} at the top of the next page. Then~\eqref{eq:outage_probability_for_singe_resource}, \eqref{eq:Q_CDF}, and~\eqref{eq:infinite_outage_prob} can, respectively, be expressed as 
\begin{align}
  P_1 \left(\gamma_{th}\right) & =  \lim_{n\to\infty}\widehat{P}_{1}\left( \mathcal{W}_n;\mathbbm{1}\right),\label{eq:p_1_limit}\\
    F_Q\left(q_{th}\right) & = \lim_{n\to\infty}\widehat{ 
  F}_Q\left( \mathcal{W}_n;\mathbbm{1}\right) ,\label{eq:q_limit_expr}\\
    P_{\infty}\left(\gamma_{th}, q_{th}\right) & = \lim_{n\to\infty} \widehat{P}_{\infty}\left( \mathcal{W}_n;\mathbbm{1}\right).\label{eq:r_limit_expr}
\end{align}
\label{thm thm2}
\end{thm}
\begin{proof} 
 See Appendix~\ref{app:A2}. \qedhere
\end{proof}

{
At this point, we are in a position to highlight the potential consequence of training an ML predictor where the focus is only on minimising 
$P_\infty \!\left( \gamma_{th}, q_{th}\right)$,
which describes the predictor's ability to minimise the outage probability of a resource given it recommends allocating that resource.
We see from \eqref{eq:r_approx} and \eqref{eq:r_limit_expr} that doing so may result in a strategy that minimises the predictor's false-negatives (the numerator of \eqref{eq:r_approx}), which can be trivially achieved by rejecting every resource. This, of course, is not desirable. Instead, considering Theorem~\ref{thm thm1}, we should also consider the number of resources available, the probabilities associated with rejecting them all, and the unconditional probability that a single resource is in outage. 
}

{
While minimising $P_\infty \!\left( \gamma_{th}, q_{th}\right)$ should not be the sole objective when training our ML predictor, it will still form a core component of the optimization problem. From~\eqref{eq:r_approx}, this would correspond approximately to minimising $\mathrm{FN} / (\mathrm{TN} + \mathrm{FN})$, which represents the proportion of negative predictions that were incorrect out of all the actual negative predictions. Importantly, this metric is not a standard one in confusion matrix analysis (e.g., precision, recall, or false-positive rate, etc,~\cite{fawcett2004roc}). Instead, it can be thought of as the proportion of all negative predictions (both true negatives and false negatives) that were actually false negatives. This gives us a measure for how ``risky" a negative prediction is. In the context of our system model, this makes sense because the negative events correspond to the ML predictor's recommendation to allocate a particular resource.
}

\section{A CUSTOM LOSS FUNCTION\label{sec:applications}}

In this section, we leverage the results that were presented in Section \ref{sec:primary_results}.B to construct a custom loss function for\footnote{{Recall that formulating a custom loss function for $\mathtt{Case~2}$ still constitutes an open problem due to the discussion preceding Theorem \ref{thm thm2}.}} {$\mathtt{Case~1}$}. We then use this loss function to train a sequence-to-sequence LSTM neural network predictor~\cite{sherstinsky2020fundamentals} in Section~\ref{sec:sims}. Because our loss function will act as an accurate approximation to $P_{\vert \mathcal{R} \vert}\left(\gamma_{th}, q_{th}\right)$ {for $\mathtt{Case~1}$} given in Theorem~\ref{thm thm1},
 we hypothesise that it will result in a trained classification predictor that is effective at minimising the system's outage probability. In our experiments that follow, this hypothesis does - indeed - appear to be true. In all scenarios tested, our novel custom loss function provides superior performance when compared to the same neural network trained with BCE, MAE, and MSE\footnote{Because the predictor is acting as a binary classifier, its output is uni-dimensional. This means that loss functions suitable for both regression and classification can be used.} loss functions {(traditional loss functions found in the literature)}. In certain scenarios, the LSTM predictor - when trained using our novel custom loss function - results in a system that achieves multiple orders of magnitude improvement in the system's outage probability. {As identified in the discussion following Theorem~\ref{thm thm1}, when the first term of \eqref{eq:main_outage_expression} and \eqref{eq:main_outage_expression_2} becomes
small (relative to the second term) we expect the two cases to
provide similar performance. As such, the custom loss function formulated for $\mathtt{Case~1}$ may be expected to provide exceptional performance for $\mathtt{Case~2}$ as well.} We now present our novel custom loss function.

\begin{figure}[t]
\vspace{-0.3cm}
\centering
\includegraphics[width=8cm]{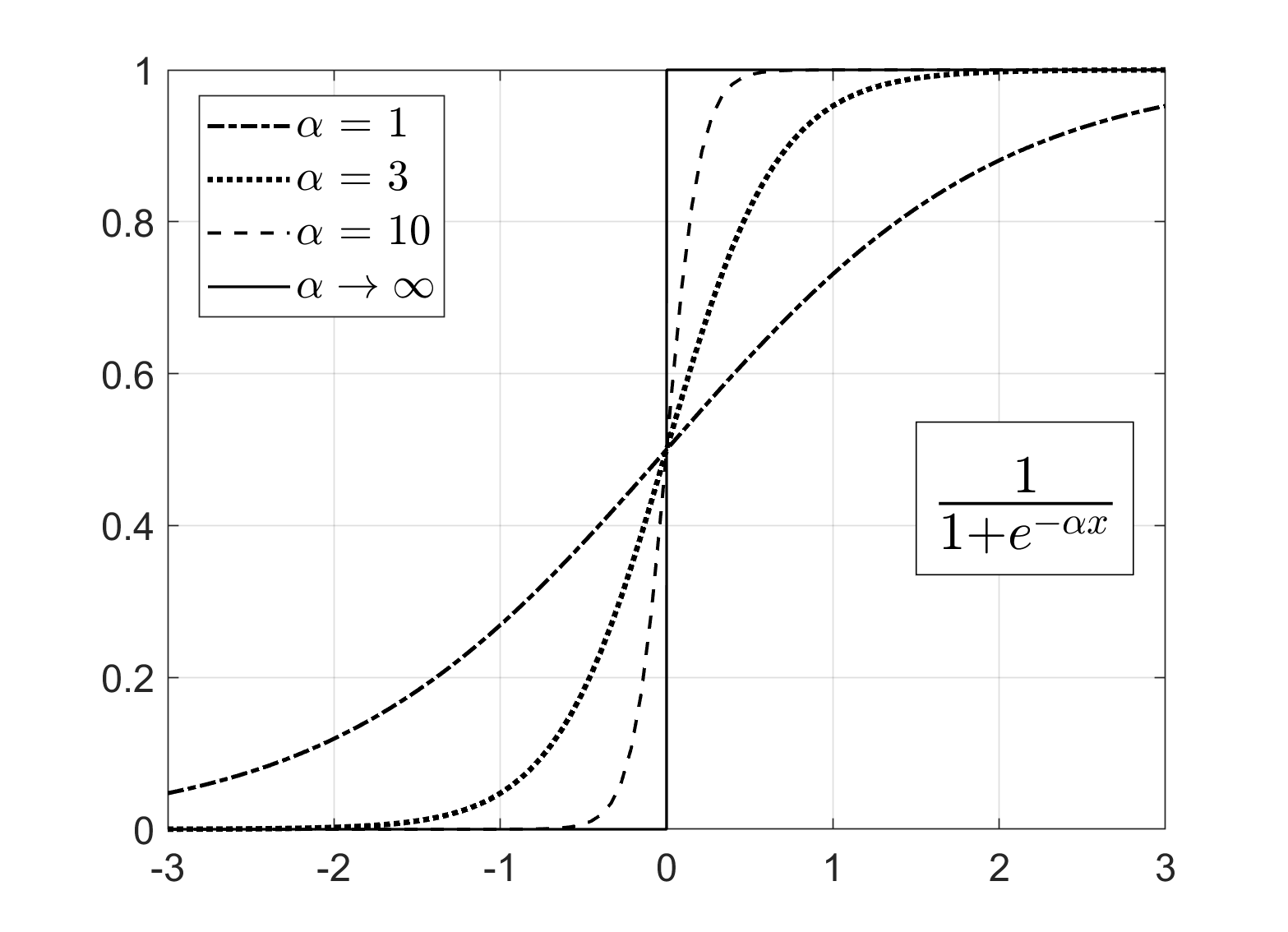}
\caption{Examples of the logistic function from \eqref{eq:logisitc}. As $\alpha\to\infty$, the function approaches the Heaviside step function.}
\label{fig:logisitc}
\end{figure}

Consider the expressions for $P_{1}\left(\gamma_{th}\right)$, $F_Q(q_{th} )$ and $P_{\infty}\left(\gamma_{th}, q_{th}\right)$ given in \eqref{eq:p_1_limit}, \eqref{eq:q_limit_expr} and \eqref{eq:r_limit_expr}, respectively. Also, consider the logistic function shown in \figref{fig:logisitc}, given by \cite{logistic}
\begin{equation}
    \phi_\alpha(x) = \frac{1}{1 + e^{-\alpha x}},\label{eq:logisitc}
\end{equation}
which forms a smooth approximation to the Heaviside step function $\mathbbm{1}(\cdot)$, becoming exact as $\alpha\to\infty$.
By dropping the limits in~\eqref{eq:p_1_limit}, \eqref{eq:q_limit_expr} and \eqref{eq:r_limit_expr}, and substituting $\mathbbm{1}(\cdot)$ with $\phi_\alpha(\cdot)$, we construct the approximations $\widehat{P}_{1}(\mathcal{W}_n;\phi_\alpha)$, $\widehat{ 
  F}_Q(\mathcal{W}_n;\phi_\alpha )$ and $ \widehat{P}_{\infty}(\mathcal{W}_n;\phi_\alpha)$. Now, using \eqref{eq:main_outage_expression} from Theorem~\ref{thm thm1} we hypothesise that minimising the following custom loss function will effectively minimise the outage probability of our system. 
\begin{definition}
With  $\widehat{P}_{1}\left(\mathcal{W}_n;\phi_\alpha\right)$, $\widehat{ 
  F}_Q\left(\mathcal{W}_n;\phi_\alpha \right)$ and $ \widehat{P}_{\infty}\left(\mathcal{W}_n;\phi_\alpha\right)$  given by \eqref{eq:p_1_approx}, \eqref{eq:q_approx} and \eqref{eq:r_approx}, respectively, the custom loss function for our system {in $\mathtt{Case~1}$} is given by
 \begin{multline}
 \ell_{\mathcal{R}}\left(\mathcal{W}_n;\phi_\alpha  \right) \triangleq~\widehat{P}_1 \left(\mathcal{W}_n;\phi_\alpha\right) \left(1-\widehat{F}_Q\left(\mathcal{W}_n; \phi_\alpha\right)\right)^{\left| \mathcal{R} \right| - 1} \\  + \widehat{P}_\infty \left(\mathcal{W}_n;\phi_\alpha\right)\left(1 - \left(1 -  \widehat{F}_Q\left(\mathcal{W}_n;\phi_\alpha\right)\right)^{\left| \mathcal{R} \right| - 1} \right). 
 \label{eq:customloss}
 \end{multline}

\label{def:customloss}
\end{definition}
\noindent{Similar} to other loss functions, our custom loss function depends on the predictor's output and labels.

\section{EXPERIMENTATION AND SIMULATION RESULTS\label{sec:sims}}

In this section, we present a collection of experimental results that apply the resource allocation strategy from Section~\ref{sec:system_model}. In these experiments, we train a lightweight LSTM neural network on simulated wireless channel data {in-line with the Clarke's 3D model~\cite{618205}} using different loss functions, including the custom loss function presented in Definition~\ref{def:customloss} above. 

\subsection{GENERATING DATA\label{sec:training_data}}
{There are various methods to generate data for any given random process. The DeepMIMO dataset, as presented in~\cite{deepmimo, DBLP:journals/corr/abs-1902-06435}, is one such example. Numerous other datasets are available in the literature.} 

{
To construct our channel data, we perform the following steps. At time $t=0$ and $\nu = 1024$, we generate a time domain vector of zero mean complex Gaussian variates with total variance $1/\nu$ given by
\begin{equation}
G\left[0\right]=\left[g_{1},g_{2},\cdots,g_{\nu}\right].
\end{equation}
To model movement in a random direction by a mobile user, at each time increment $t$, we apply a per-element independent and uniformly distributed phase shift $\theta_{it}\sim \mathrm{Unif}\left[ -\zeta , \zeta \right]$.
At time $t=1$, we have
\begin{equation}
G\left[1\right]=\left[e^{i\theta_{11}}g_{1},e^{i\theta_{21}}g_{2},\cdots,e^{i\theta_{\nu1}}g_{\nu}\right],
\end{equation}
and so on. In the frequency domain we, respectively, have
\begin{equation}
{\mathcal{H}}\left[0\right]=\frac{1}{\sqrt{\nu}}\left[\sum_{m=0}^{\nu-1}g_{m}e^{-i2\pi m/\nu}, \cdots,\sum_{m=0}^{\nu-1}g_{m}e^{-i2\pi m}\right]
\end{equation}
and
\begin{equation}
{\mathcal{H}}\left[1\right]\!=\!\frac{1}{\sqrt{\nu}}\!\left[\!\sum_{m=0}^{\nu-1}\!\!e^{i\theta_{m1}}g_{m}e^{-i2\pi m/\nu},\cdots,\!\!\sum_{m=0}^{\nu-1}\!\!e^{i\theta_{m1}}g_{m}e^{-i2\pi m}\!\right]\!.
\end{equation}
Letting $h_i\left[t\right]$ denote the $i$th element of ${\mathcal{H}}\left[t\right]$, we model the channel for resource $i$ at time $t$ as the narrowband system
\begin{equation}
y_{i}\left[t\right]=h_i\left[t\right] x_{i}\left[t\right]+w_{i}\left[t\right],
\end{equation}
where $x_{i}\left[t\right]$ is a unit variance signal term and  $w_{i}\left[t\right]$ is a unit variance zero mean complex Gaussian noise term, which yields an average SNR of $$\mathtt{SNR} = \mathbb{E}\left[\left\Vert h_i\left[t\right]\right\Vert^2\right],$$
where $\mathbb{E}\left[\cdot\right]$ is the expectation operator.
Clearly, the variables $h_i\left[t\right]$ are stationary and zero mean complex Gaussian. Also, the autocorrelation between successive channel samples is given by:
\begin{align}
\mathbb{E}\left[h_i\left[t\right] h_i\left[t+1\right] \right] &  =\mathbb{E}\left[\frac{1}{\nu}\sum_{m=0}^{\nu-1}e^{i\theta_{m1}}\left\Vert g_{m}\right\Vert ^{2}\right] \nonumber \\
 & =\mathbb{E}\left[\left\Vert g_{m}\right\Vert ^{2}\right]\mathbb{E}\left[e^{i\theta_{11}}\right] \nonumber\\
 & =\mathbb{E}\left[\left\Vert g_{m}\right\Vert ^{2}\right]\mathrm{sinc}\left( \zeta\right),
\end{align}
where the final line follows from the characteristic function of the uniform distribution~\cite[Eq. (5)]{uniform_wolfram}. This model is in-line with the 3D Clarke's model presented in~\cite[Eq. (7)]{618205}, where $\zeta$ corresponds to the straight-line distance travelled by the mobile user. This model\footnote{{Additionally, channel data in-line with the Clarke's 2D model~\cite{6779222} was generated to validate the efficacy of our loss function. It should be noted that the results to be discussed in the subsequent subsections are equally applicable to this dataset.}} is widely accepted by the wireless community.}

We use the above method to generate a sequence of $k+l$ channel samples for each resource. In our experiments, we consider $k = 100$ and $l=10$. We also set $ \zeta = 0.1$ radians. The first $k$ samples {$H_i(t,k)$} are used as input to our LSTM neural network. The final $l$ samples {$H_i(t+l,l)$} are used to construct a label $b_{i}$, identifying whether communication could have been supported at a particular rate. {Specifically, using the formula from \eqref{eq:Gaussian_channel_capacity} to determine a capacity, we have:
\begin{align}
b_i = \begin{cases} 
      0 & C\!\left(H_i(t+l,l)\right) \geq \gamma_{th} \\
      1 & C\!\left(H_i(t+l,l)\right)  < \gamma_{th}.
   \end{cases}
\end{align}}

\subsection{TRAINING}


We employ a lightweight LSTM neural network predictor consisting of a single LSTM layer with 32 hidden units. {Subsequent to this layer are two dense layers: the first contains 10 units activated by PReLU, while the second has a singular unit activated by a sigmoid function. Our model is trained on a dataset comprising (4,500 x number of resources) samples. This means that for a 4 resource system, our model is trained on a dataset consisting of 18,000 samples. For an 8 resource system, this would be 36,000 samples, and so on.} This dataset, {${\mathcal{W}}_n$}, is generated as described in Section V.\ref{sec:training_data}. {Furthermore, the size of the validation dataset matches that of the training dataset and we test on 13,000 instantiations of the $\vert{\mathcal{R}}\vert$ resource system. This number is chosen to ensure statistically robust results. Additionally, we retrain the predictor 10 times, taking the average of the performance to obtain a single data point in each of the figures. This has the effect of performing the complete test on (130000 x number of resources) samples. } 
For the training, {we adopt a supervised learning approach}, leverage TensorFlow's Keras API, utilize an ADAM optimizer, and employ the following loss functions: BCE, MAE, MSE, and the custom loss function from Definition~\ref{def:customloss}.

\begin{figure}[t]
\vspace{-0.3cm}
\centering
\includegraphics[width=8.6cm]{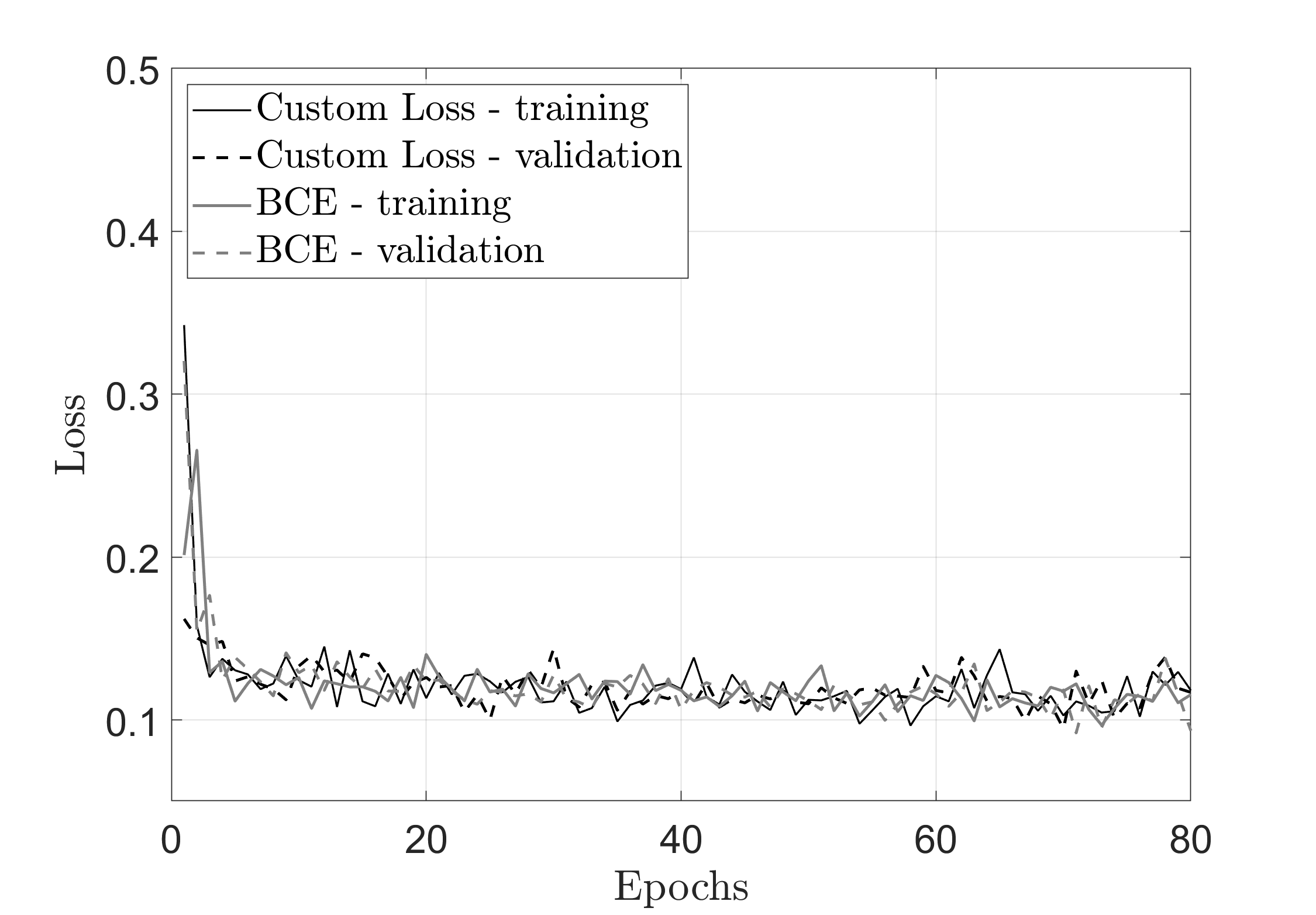}
\vspace{-0.5cm}
\caption{Training versus Validation loss curve generated for a 10 resource system with rate threshold $\gamma_{th}=0.5$, {$\mathtt{SNR} = 0~\rm{dB}$} and $q_{th}=0.5$.}
\label{fig:loss_Curve}
\end{figure}

{The key hyperparameters that were chosen to train our LSTM model are provided in Table~\ref{table:hyperparameters}.   
Additionally, we employed the early stopping technique during training to avoid unnecessarily long training routines and mitigate overfitting, setting a threshold of 30 epochs based on observed performance metrics. This decision was substantiated by the convergence patterns of the training and validation loss curves, which maintained close alignment throughout the training period, as depicted in~\figref{fig:loss_Curve}. This figure also makes it clear that our model does not display either overfitting or under-fitting behaviour.
Finally, it should also be noted that an independent test set, was used to validate all subsequent results, ensuring they were not influenced by any potential model overfitting.}

\begin{table}[t]
    \centering
    \renewcommand{\arraystretch}{0.9}
    \caption{Model architecture and chosen hyperparameters for the LSTM training and custom loss function.}
    \begin{tabular}{|l|l|}
    \hline
    \textbf{Architecture} & \textbf{Specification} \\
    \hline
    Model type & Lightweight LSTM \\
    Layer 1: LSTM layer & 1 layer; 32 hidden units \\
    Layer 2: Dense layer & 10 units with PReLU activation \\
    Layer 3: Dense layer & 1 unit with sigmoid activation \\
    \hline
    \textbf{Hyperparameter} & \textbf{Value} \\
    \hline
    Learning rate & 0.001 \\
    Number of epochs & 30 \\
    Epoch size & 150 \\
    Batch size & Number of resources \\
    Input sequence length & 100 \\
    Output sequence length & 10 \\
    \hline
    \textbf{Hyperparameters: Custom Loss} & \textbf{Value} \\
     \hline
    $\alpha$ & 10 \\
    $q_{th}$ & 0.5 \\
    \hline
    \end{tabular}
   \label{table:hyperparameters}
\end{table} 

For all of our results, our predictor with the custom loss function\footnote{For the other loss functions, $q_{th}$ is not a hyperparameter of the training procedure.} is trained with a prediction classification threshold of $q_{th}=0.5$ and {$\alpha = 10$}. For a fixed rate threshold and fixed number of resources, a single predictor took approximately 5~minutes to train on a desktop computer using an NVIDIA\textsuperscript{\textregistered} GeForce\textsuperscript{\textregistered} RTX 2080~Ti 11~GB GDDR6 GPU.  The code developed that supports our work can be found~\href{https://github.com/ML4Comms/greedy-resource-allocation-outage-classification}{here}~\cite{custom_loss}.

\begin{figure}[t]
\centering
\vspace{-1.5cm}
	\begin{minipage}[c][1\width]{
	   0.48\textwidth}
	   \centering
	   \includegraphics[width=1.0\textwidth]{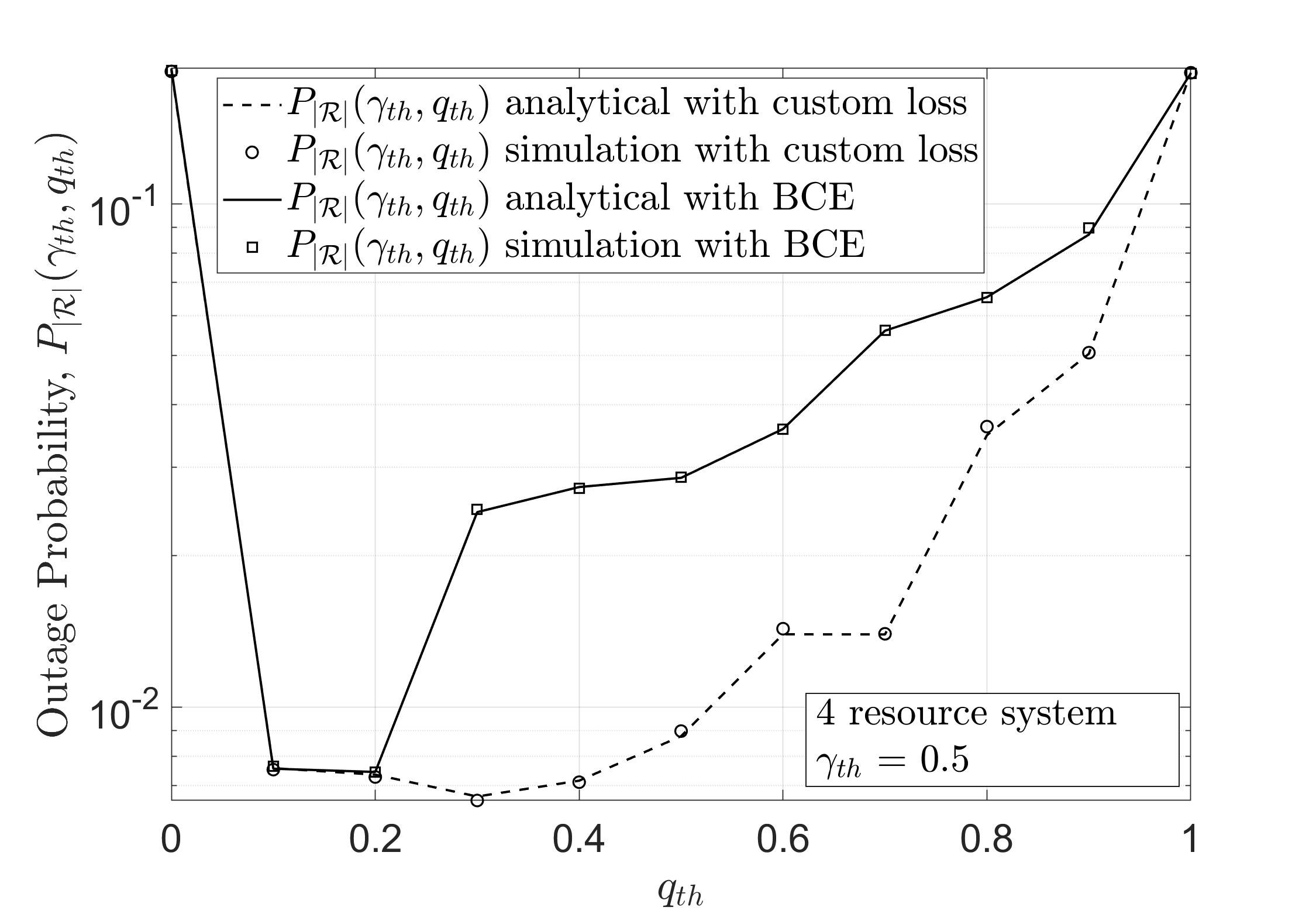}
	\end{minipage} 
	\vspace{-2.6cm} \\
	\begin{minipage}[c][1\width]{
	   0.48\textwidth}
	   \centering
	   \includegraphics[width=1.0\textwidth]{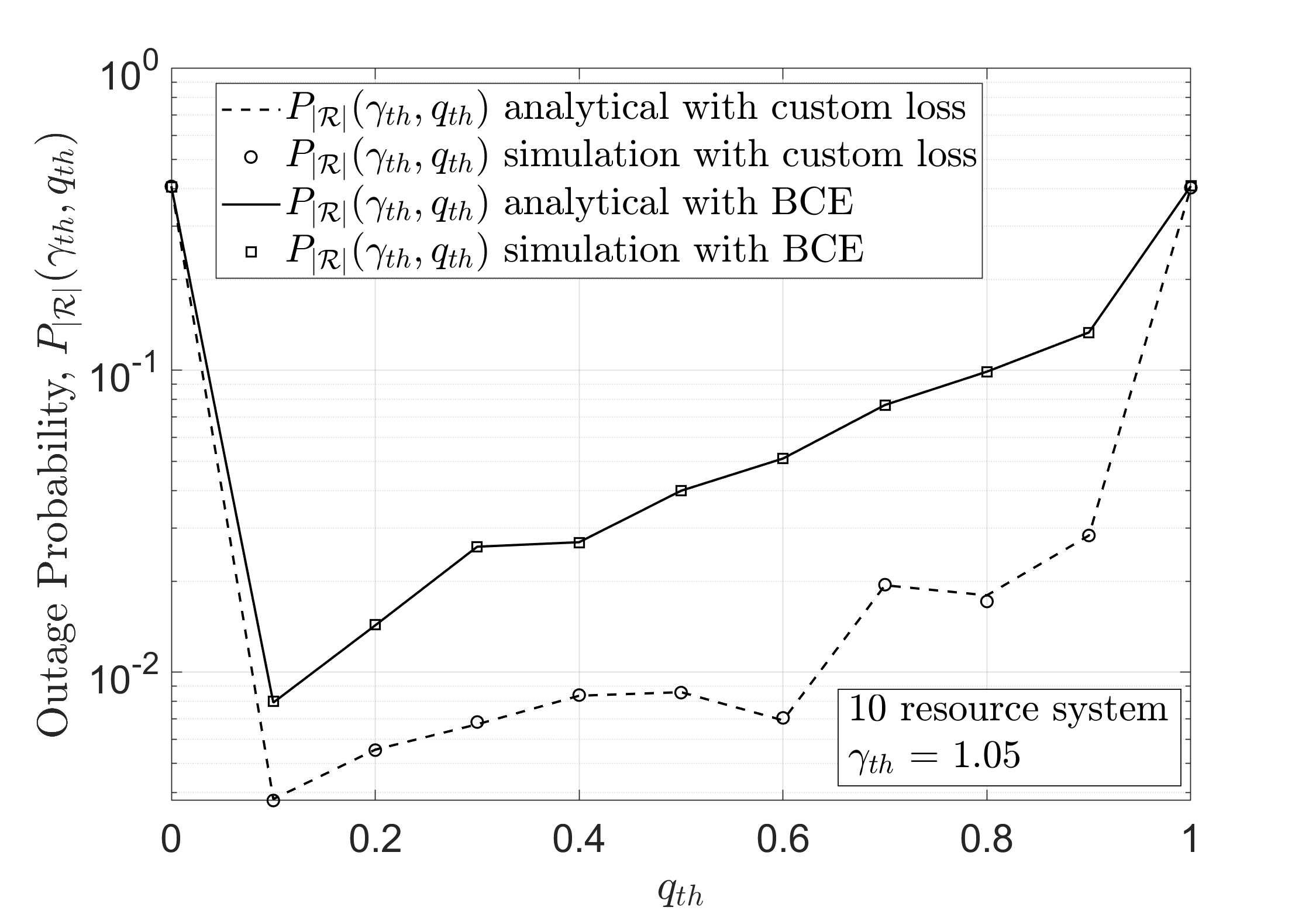}
	\end{minipage}
 \vspace{-1.2cm}
\caption{Analytical and Monte Carlo simulation results for the outage
probability of the system $P_{\vert \mathcal{R} \vert} \left(\gamma_{th}, q_{th}\right)$ {in~$\mathtt{Case~1}$}~\eqref{eq:main_outage_expression}, using the custom \eqref{eq:customloss} or
BCE loss function. Each figure considers a different number of available
resources $\vert \mathcal{R} \vert$ and rate thresholds with {$\mathtt{SNR} = 0~\rm{dB}$. For the custom loss function the model was trained with $q_{th} = 0.5$.}}
\label{fig:OP_vs_qth}
\end{figure}



\begin{figure}[t]
\vspace{-0.3cm}
\centering
\includegraphics[width=8.6cm]{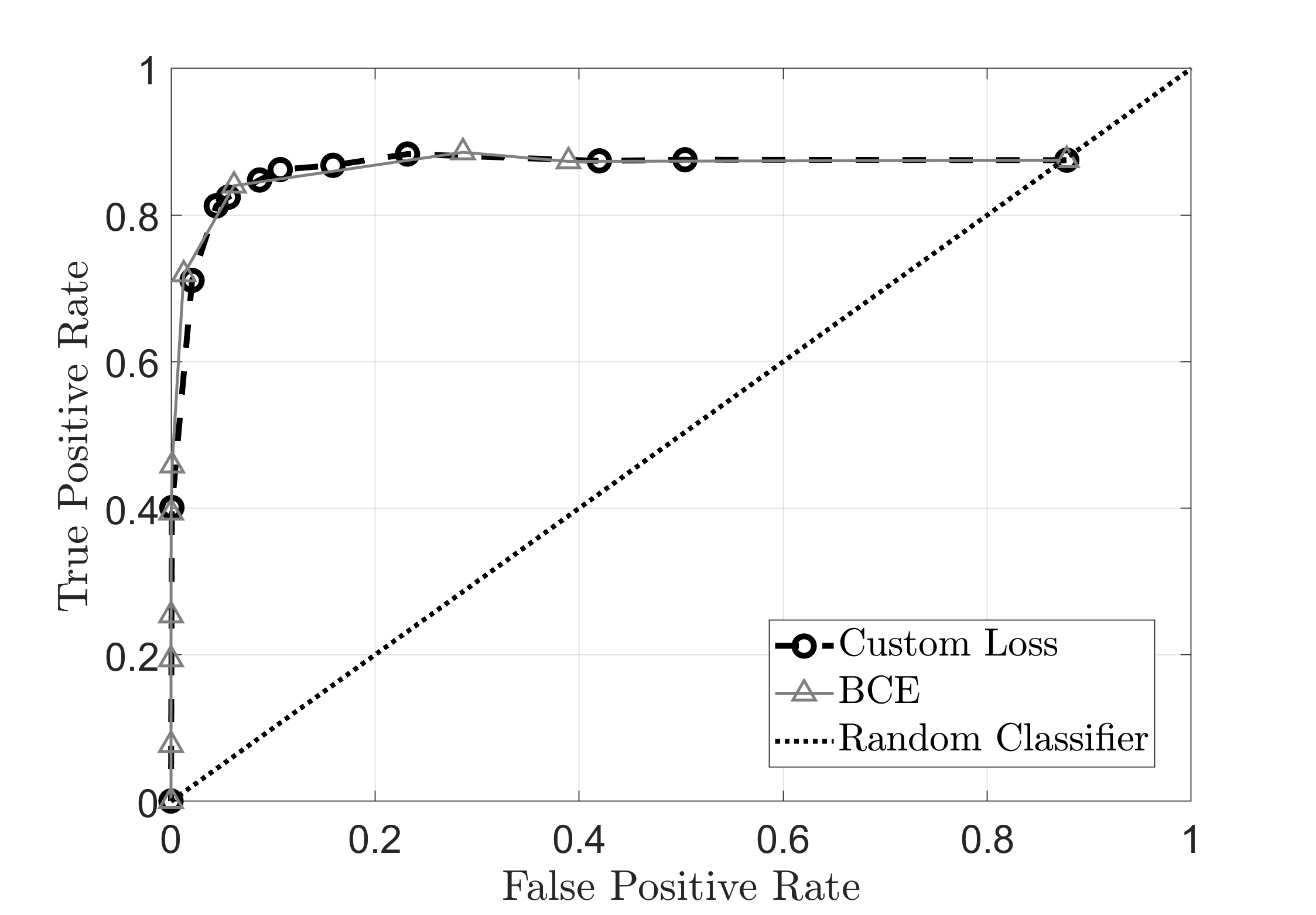}
\vspace{-0.5cm}
\caption{{ROC curve generated for a 10 resource system with rate threshold $\gamma_{th}=0.5$ and $\mathtt{SNR} = 0~\rm{dB}$. Each point is parametrically generated by sweeping $q_{th}$ between 0 and 1.}}
\label{fig:roc}
\end{figure}

\subsection{RESULTS}

\begin{figure}[t]
\centering
\vspace{-1.5cm}
	\begin{minipage}[c][1\width]{
	   0.48\textwidth}
	   \centering
	   \includegraphics[width=1.0\textwidth]{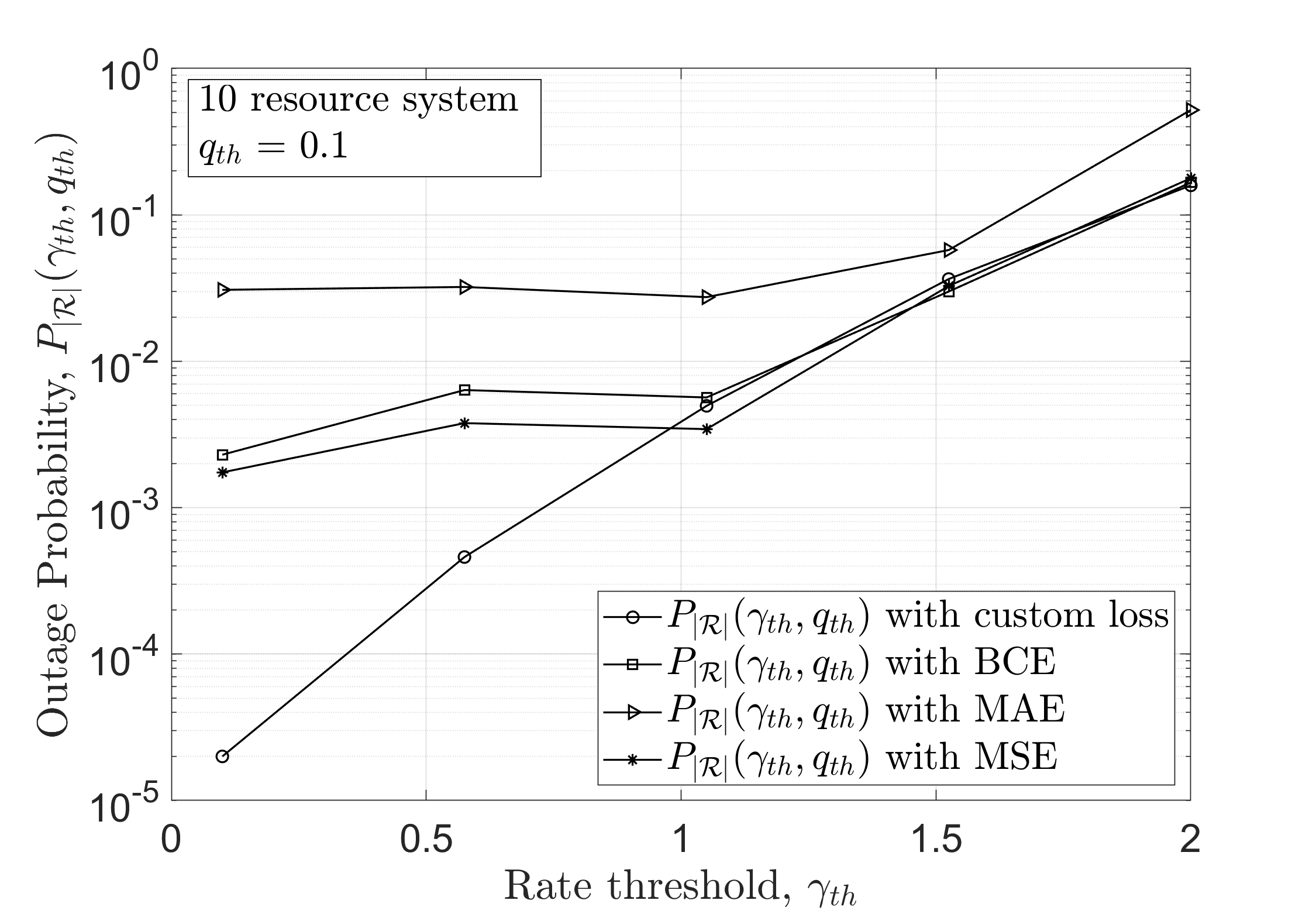}
	\end{minipage} 
	\vspace{-2.6cm} \\
	\begin{minipage}[c][1\width]{
	   0.48\textwidth}
	   \centering
	   \includegraphics[width=1.0\textwidth]{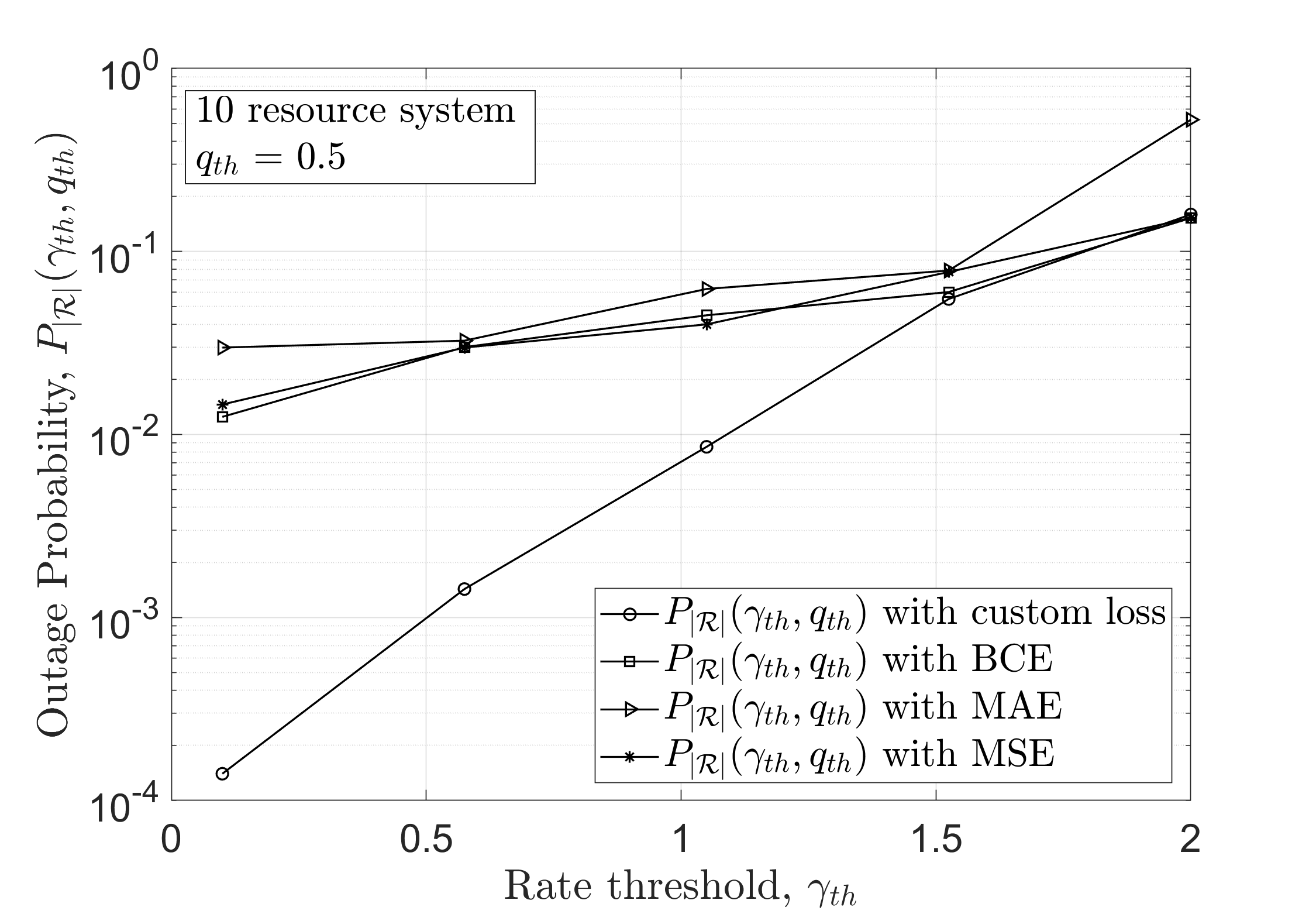}
	\end{minipage}
 \vspace{-1.2cm}
\caption{Comparison of Monte Carlo simulation results for the outage probability of the system $P_{\vert \mathcal{R} \vert} \left(\gamma_{th}, q_{th}\right)$ {in~$\mathtt{Case~i}$}~\eqref{eq:main_outage_expression} with 10 resources for different rate thresholds with {$\mathtt{SNR} = 0~\rm{dB}$} when using the custom \eqref{eq:customloss}, BCE, MAE and MSE loss functions. Each figure considers a different $q_{th}$.}
\label{fig:OP_vs_rate}
\end{figure}
{As mentioned in the previous subsection, in every figure that follows, each data point is created by retraining the predictor 10 times and taking the average performance of the model. Moreover, during each of these iterations, the model evaluated its performance over 13,000 instantiations of the $\vert\mathcal{R}\vert$ resource system, a number chosen to ensure statistically robust results.}
\begin{figure}[t]
\centering
\vspace{-2.1cm}
	\begin{minipage}[c][1.1\width]{
	   0.48\textwidth}
	   \centering
	   \includegraphics[width=1.0\textwidth]{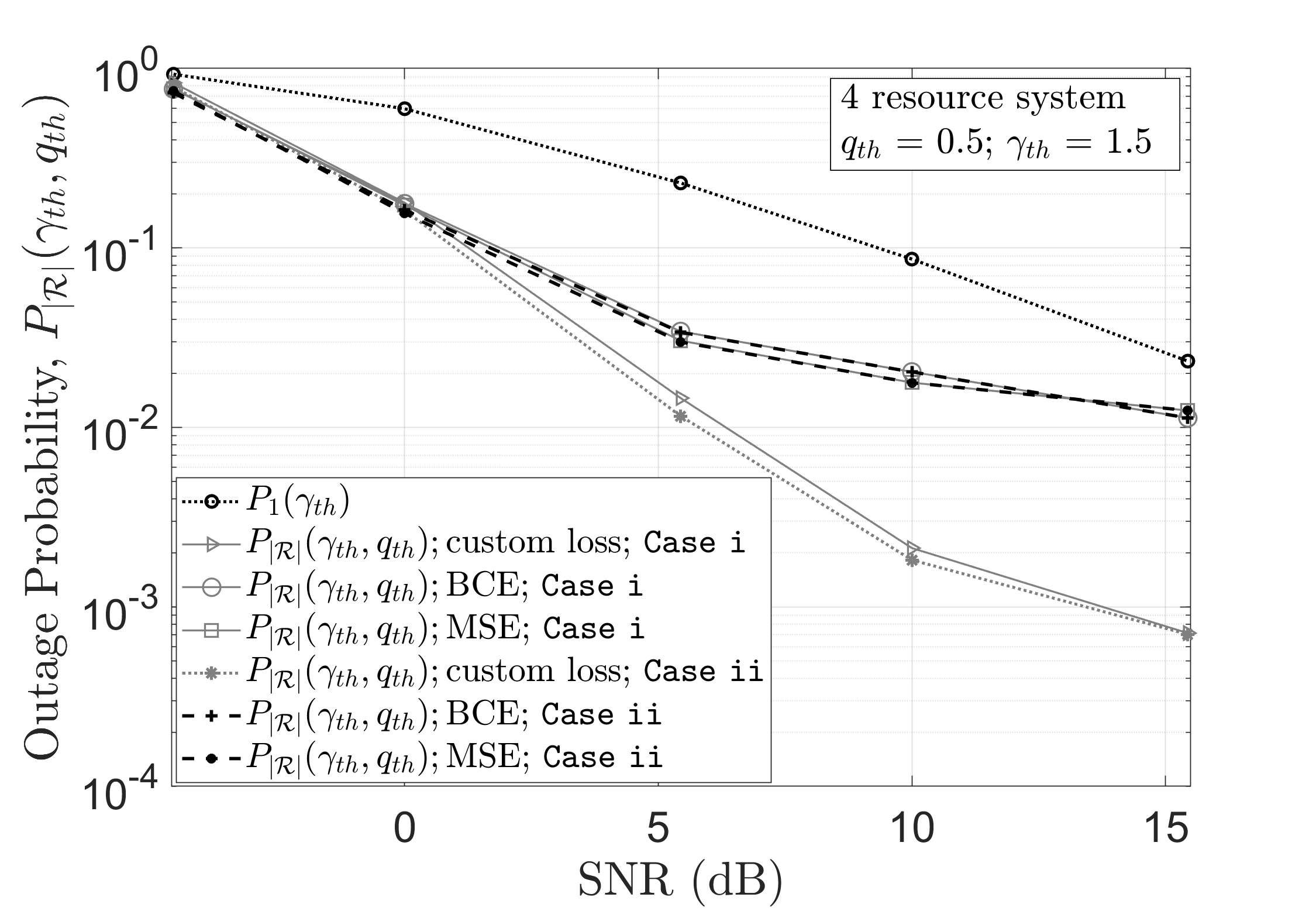}
	\end{minipage} 
	\vspace{-3.1cm} \\
	\begin{minipage}[c][1\width]{
	   0.48\textwidth}
	   \centering
	   \includegraphics[width=1.0\textwidth]{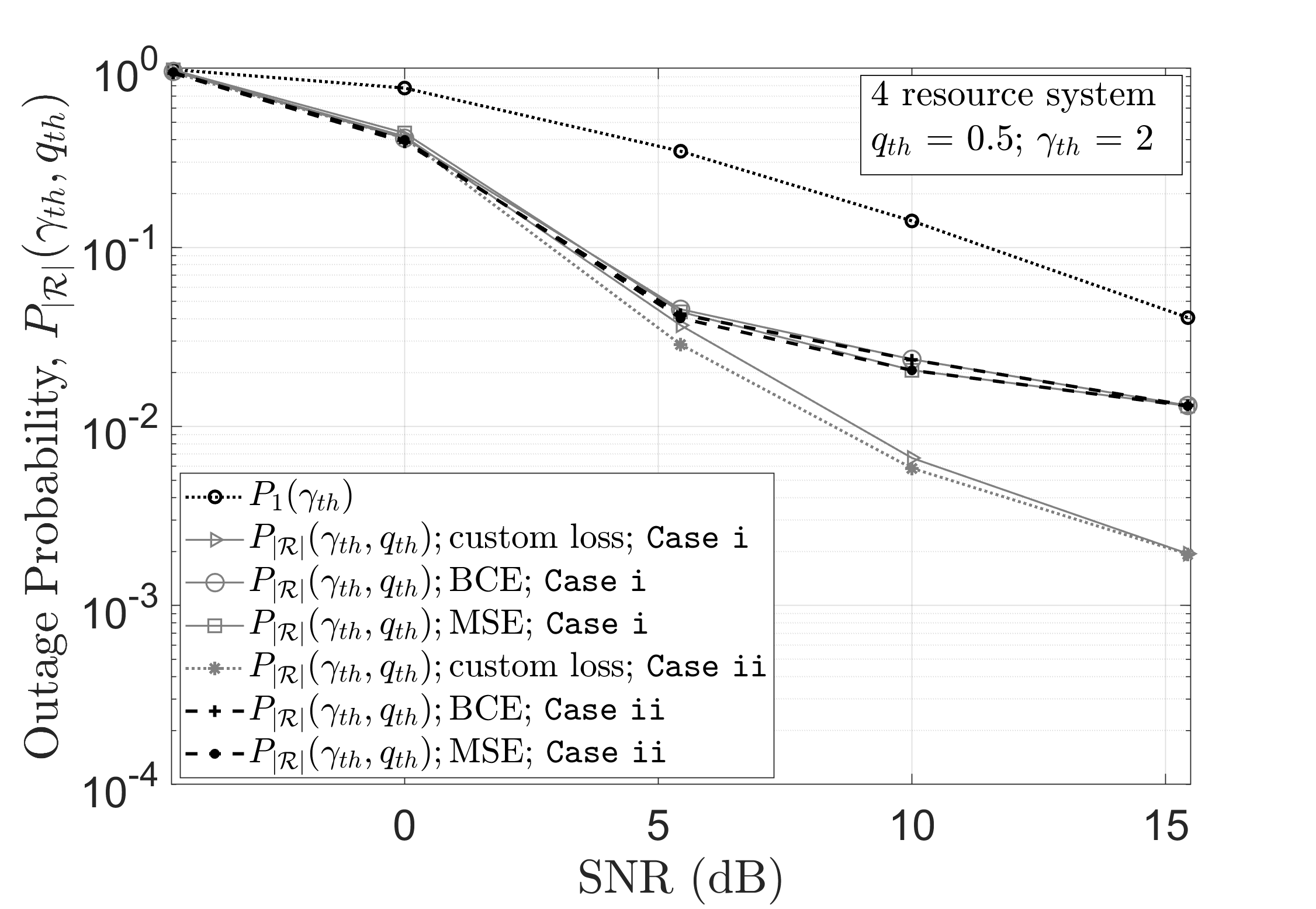}
	\end{minipage}
 \vspace{-1.2cm}
\caption{{Comparison of Monte Carlo simulation results for the outage probability of the system $P_{\vert \mathcal{R} \vert} \left(\gamma_{th}, q_{th}\right)$ in $\mathtt{Case~1}$~\eqref{eq:main_outage_expression} and~$\mathtt{Case~2}$~\eqref{eq:main_outage_expression_2} with 4 resources for different SNRs when using the custom \eqref{eq:customloss}, BCE, and MSE loss functions. Each figure considers a different $\gamma_{th}$. $P_1\left(\gamma_{th}\right)$ given in~\eqref{eq:outage_probability_for_singe_resource}, is also plotted for comparison and is the same for all loss functions.}}
\label{fig:OP_vs_SNR_4R}
\end{figure}

Fig.~\ref{fig:OP_vs_qth} shows the outage performance of {a 4 and 10 resource system for $\mathtt{Case~1}$} when using different loss functions and rate thresholds, $\gamma_{th}$, for a range of predictor classification thresholds, $q_{th}$. We would expect to see a `U' or `V' shaped curve, indicating that too small or too large values of $q_{th}$ are not good, whilst the best value lies somewhere in between. Since the predictor was trained at $q_{th} = 0.5$, a reasonably low outage probability is observed at this value. 
At the limiting values of $q_{th} = 0$ and $q_{th} = 1$, we observe that the outage probability of the system equals that of a single resource system, a behaviour that was suggested in the discussion following Theorem~\ref{thm thm1}.  It can also be seen that the range of $q_{th}$ values over which the custom loss function performs reasonably well is significantly broader than that of the BCE loss function.
Furthermore, at $q_{th} = 0$ and $q_{th} = 1$, we can see that the predictor's performance for different loss functions coincide. This is because $q_{th}=0$ or $1$ will, respectively, result in the user always being allocated to the final or first resource, irrespective of the predictors output. 


{
As highlighted in the discussion following Theorem~\ref{thm thm2}, a core component of training our ML predictor is to minimise $\mathrm{FN}/(\mathrm{TN}+\mathrm{FN})$. Importantly, this is not the same as optimising for recall (sometimes called true-positive rate) ($\mathrm{TP}/(\mathrm{TP}+\mathrm{TN})$),  false positive rate ($\mathrm{FP}/(\mathrm{FP}+\mathrm{TN})$), or precision ($\mathrm{TP}/(\mathrm{TP}+\mathrm{FP})$), etc, of our predictor. As such, we may not expect our ML predictor with our custom training procedure to provide standout performance with respect to these measures. It is, however, of interest to observe how it  performs with respect to these more standard metrics. Fig.~\ref{fig:roc} shows an ROC curve~\cite{fawcett2004roc} for a predictor trained using our custom loss function or BCE. Despite not having optimised for either true-positive rate or false-positive rate, our ML predictor still performs well, providing on-par performance with  BCE. Similar conclusions are found for other configurations too.
}

Fig.~\ref{fig:OP_vs_rate} shows the outage performance of a $10$ resource system {for $\mathtt{Case~1}$} when using different loss functions and different predictor classification thresholds $q_{th}$, for a range of rate thresholds $\gamma_{th}$. It is clear that the custom loss function formulated in this paper provides similar or superior performance when compared to the others. For regions where outages occur infrequently, i.e., for low rate thresholds, our novel loss function becomes increasingly more dominant. As an example, in both figures, our custom loss function may improve the performance of the system by approximately two orders of magnitude. In the same $\gamma_{th}$ region, the BCE, MAE, and MSE loss functions perform poorly which is most likely due to the increasing bias in the dataset as outages become more rare {(i.e., as outage events go deeper into the tail of the channel's distribution)}. Notably, without making \textit{any} corrections for this bias (i.e., imbalance in the proportion of outages to non-outages), our custom loss function is able to improve the system's performance dramatically, {thus demonstrating deep-tail learning.}
{Deep-tail learning is significant because it enables our model to accurately capture the extremely low probability tail-end of the channel's distribution, which is where outage events are present for low values of the rate threshold $\gamma_{th}$.}

\begin{figure}[t]
\vspace{-0.3cm}
\centering
\includegraphics[width=8.6cm]{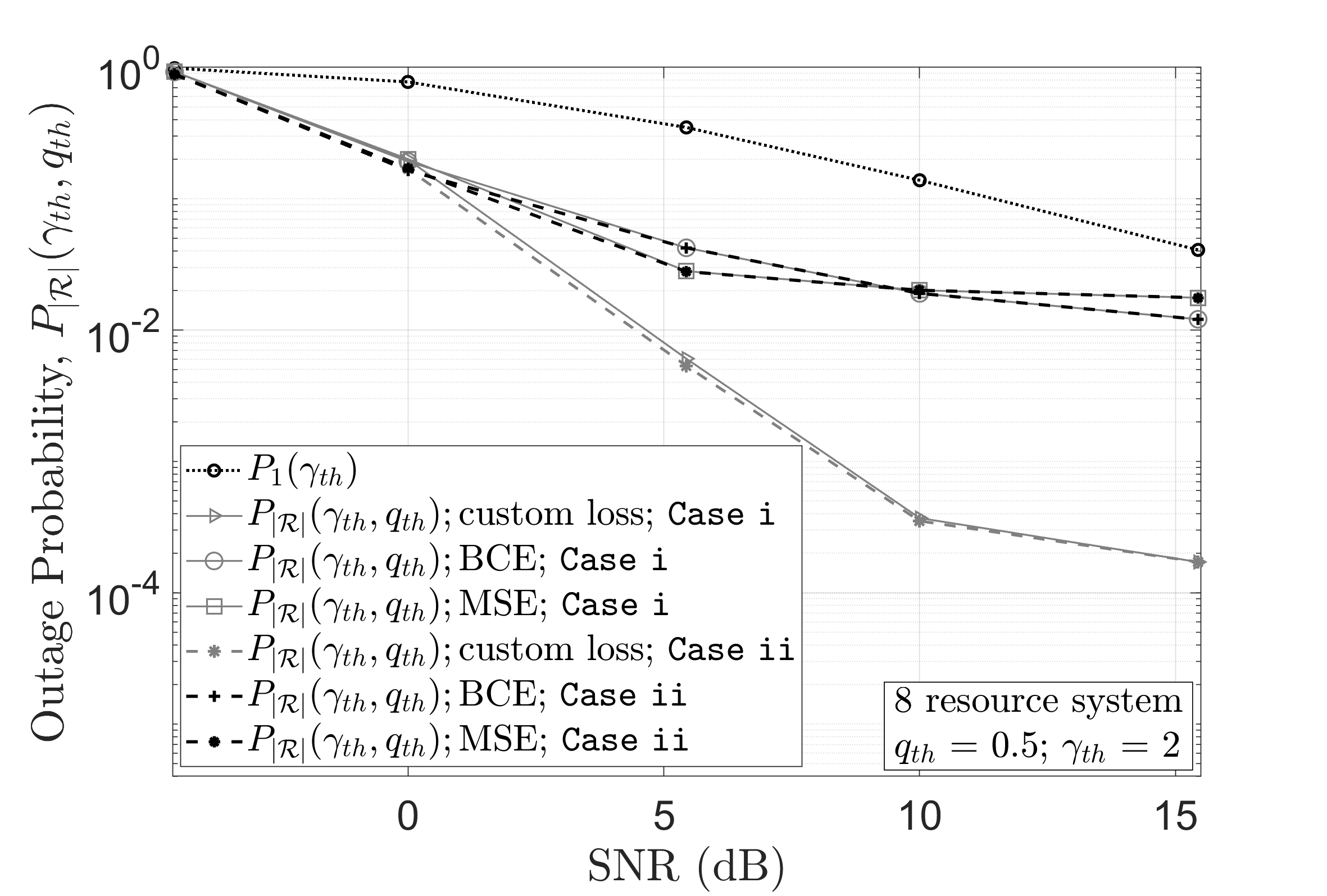}
\vspace{-0.5cm}
\caption{{Comparison of Monte Carlo simulation results for the outage probability of the system $P_{\vert \mathcal{R} \vert} \left(\gamma_{th}, q_{th}\right)$ in $\mathtt{Case~1}$~\eqref{eq:main_outage_expression} and~$\mathtt{Case~2}$~\eqref{eq:main_outage_expression_2} with 8 resources for different SNRs when using the custom \eqref{eq:customloss}, BCE, and MSE loss functions. Here $\gamma_{th} = 2$. $P_1\left(\gamma_{th}\right)$ given in~\eqref{eq:outage_probability_for_singe_resource}, is also plotted for comparison and is the same for all loss functions.}}
\label{fig:OP_vs_SNR_8R}
\end{figure}

\begin{figure}[t]
\vspace{-0.3cm}
\centering
\includegraphics[width=8.6cm]{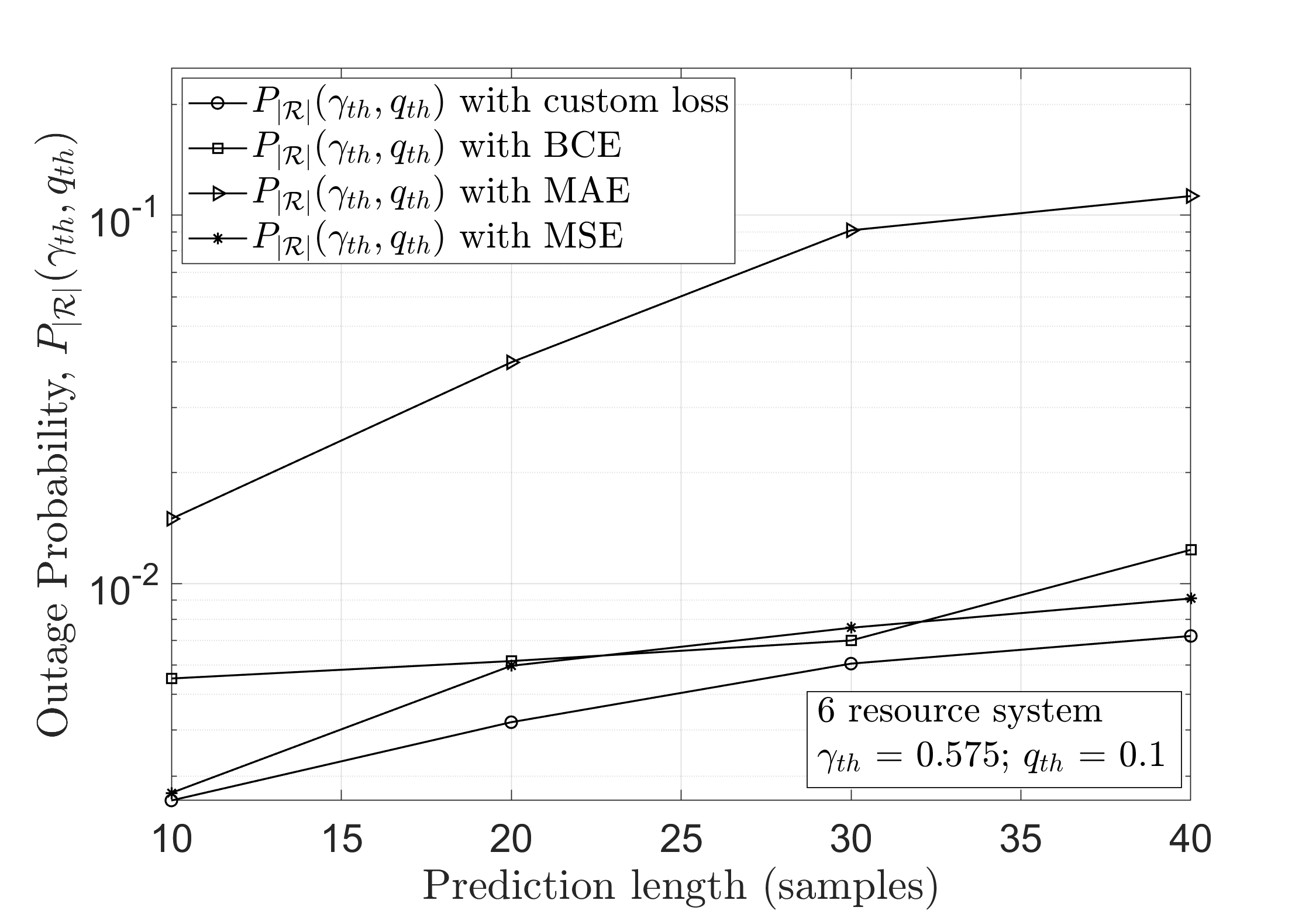}
\vspace{-0.5cm}
\caption{Comparison of Monte Carlo simulation results for the outage probability of the system $P_{\vert \mathcal{R} \vert} \left(\gamma_{th}, q_{th}\right)$ {in $\mathtt{Case~1}$}~ \eqref{eq:main_outage_expression} for different prediction lengths when using the custom \eqref{eq:customloss}, BCE, MAE and MSE loss functions. We consider a $6$ resource system, $q_{th} = 0.1$, $\gamma_{th} = 0.575$ and {$\mathtt{SNR} = 0~\rm{dB}$.}}
\label{fig:OP_vs_predictionlength}
\end{figure}
{Figs.~\ref{fig:OP_vs_SNR_4R} and~\ref{fig:OP_vs_SNR_8R} show the outage performance of a 4 and 8 resource system, respectively, for cases~1 and 2 when using the BCE, MSE and custom loss functions for a range of $\mathtt{SNR}$ and two different rate thresholds. The outage probability of a single resource system i.e., $P_1\left(\gamma_{th}\right)$ in~\eqref{eq:outage_probability_for_singe_resource} is also plotted for comparison and is the same for all loss functions. As before, the custom loss function developed in this work demonstrates comparable or enhanced performance relative to other loss functions, showing strong dominance in high $\mathtt{SNR}$ regions.
It is also evident that the custom loss function that was specifically formulated for $\mathtt{Case~1}$ performs  almost identically for $\mathtt{Case~2}$.} 

Fig.~\ref{fig:OP_vs_predictionlength} shows the outage performance of a 6 resource system for {$\mathtt{Case~1}$} when using different loss functions, $q_{th} = 0.1$ and $\gamma_{th} = 0.575$, for a range of output prediction lengths. Again, the custom loss function performs similarly or substantially improves upon the performance attained using the other loss functions.

\section{CONCLUSIONS\label{sec:conc}}

In this work, novel expressions were presented for the outage probability of a single-user multi-resource allocation system that exploits an ML binary classification predictor to predict and avoid future resource outages.  
 {Through these expressions, we revealed that optimization of the per-resource outage probability conditioned on the ML predictor recommending resource allocation may produce inadequate predictors that reject every resource. Also, focusing on standard metrics, like precision, false-positive rate, or recall, may again produce inadequate predictors.}
We then proposed a novel custom loss function, which approximates the theoretical outage probability, is differentiable, and can be calculated using our dataset. Importantly, it was demonstrated that predictors trained with our novel loss function provided exceptionally competitive performance when compared to those trained with BCE, MAE or MSE loss functions. For example, in some cases we observed performance improvements of approximately two orders of magnitude over all other tested loss functions. Moreover, this improvement was notable even without addressing the imbalance in the proportion of outages to non-outages in the dataset, {showcasing deep-tail learning}.

\section*{APPENDIX}

	\subsection{PROOF OF THEOREM~\ref{thm thm1}  $\label{app:A0}$}

\subsubsection*{{$\mathtt{Case~1}$:}}

To prove this theorem {for $\mathtt{Case~1}$,} we define the resource allocation event $\mathtt{E}_{i}$, {shown in \eqref{eq:theorem_1_union} at the top of the next page.} This takes two forms. 
The first is when $i < \left|\mathcal{R}\right|$, i.e., when the number of available resources to select from is greater than 0. The second is when $i = \left|\mathcal{R}\right|$, i.e., when there are no extra resources available, in which case the predictor's prediction is ignored.
\begin{figure*}
{
    \begin{align}
    \mathtt{E}_{i}&=\begin{cases} 
         ~~\bigcap_{i'=1}^{i-1}\left\{Q\left(H_{i'}{(t,k)}\right) > q_{th} \right\}~\bigcap~ \left\{Q(H_{i}{(t,k)}) \leq q_{th} \right\} & ~~i<|\mathcal{R}|\\
         & \\
          ~~\bigcap_{i'=1}^{|\mathcal{R}|-1}\left\{Q\left(H_{i'}{(t,k)}\right) > q_{th} \right\} &~~ i=|\mathcal{R}|
       \end{cases}\label{eq:theorem_1_union}
    \end{align}
    \hrulefill
}
\end{figure*}

This allows us to define the following event that corresponds to the simultaneous occurrence of resource outage and resource allocation:
{
\begin{equation}
\widehat{\mathtt{E}}_{i} \triangleq \left\{C({H}_{i}\left(t+l, l\right)) < \gamma_{th}\right\}\cap {\mathtt{E}}_{i}.\label{eq:Ehat_r}
\end{equation}
}
Thus, for the greedy resource allocation strategy, the outage probability of the system~\eqref{eq:main_outage_expression} is given by the probability of the union over all possible events, $\widehat{\mathtt{E}}_{i}$: 
\begin{equation}
P_{\vert \mathcal{R} \vert}\left(\gamma_{th},q_{th}\right) = \mathbb{P} \left[  \bigcup_{i=1}^{\left|\mathcal{R}\right|} \widehat{\mathtt{E}}_{i}  \right].\label{eq:R_outage_def}
\end{equation}

We now prove \eqref{eq:infinite_outage_prob}, the outage probability for an infinite resource system. For this, see~\eqref{eq:infiniteproof1},~\eqref{eq:infiniteproof2} and~\eqref{eq:infiniteproof3}, shown at the top of the next page.
\begin{figure*}[ht]
\begin{align}
P_{\infty}\left(\gamma_{th},q_{th}\right)   = &~\mathbb{P}\left[\bigcup_{i=1}^{\infty}\widehat{\mathtt{E}}_{i} \right]=\sum_{i=1}^{\infty}\mathbb{P}\left(\widehat{\mathtt{E}}_{i}\right) \label{eq:infiniteproof1} \\
=&\sum_{i=1}^{\infty}\mathbb{P}\left[C\left(H_{i}{(t+l,l)}\right) < \gamma_{th} \bigcap_{i'=1}^{i-1}Q\left(H_{i'}{(t,k)}\right) > q_{th}\bigcap Q\left(H_{i}{(t,k)}\right) \leq q_{th}\right] \label{eq:infiniteproof2} \\
=&\sum_{i=1}^{\infty}\mathbb{P}\left[ C\left(H_{i}{(t+l,l)}\right) < \gamma_{th}~\big\vert \bigcap_{i'=1}^{i-1}Q(H_{i'}{(t,k)}) > q_{th}\bigcap Q\left(H_{i}{(t,k)}\right) \leq q_{th} \right] 
\nonumber \\ &\times
\mathbb{P}\left[\bigcap_{i'=1}^{i-1}Q\left(H_{i'}{(t,k)}\right) > q_{th}\bigcap Q\left(H_{i}{(t,k)}\right) \leq q_{th}\right]. 
\label{eq:infiniteproof3}
\end{align}
\hrulefill
\end{figure*}
Here,~\eqref{eq:infiniteproof1} follows from $\widehat{\mathtt{E}}_{i}$ and $\widehat{\mathtt{E}}_{i'}$ being mutually exclusive\footnote{$\widehat{\mathtt{E}}_{i}$ and $\widehat{\mathtt{E}}_{i'}$ are mutually exclusive because the system does not allow the selection of resources $i$ and $i'$ simultaneously for $i\neq i'$.} for $i \neq i'$, \eqref{eq:infiniteproof2} follows from \eqref{eq:Ehat_r}, and~\eqref{eq:infiniteproof3} follows from Bayes' theorem. Since ${H}_{i}\left(\cdot, \cdot \right)$ and ${H}_{i'}\left(\cdot, \cdot \right)$ are independent for $i \neq i'$, \eqref{eq:infiniteproof3} simplifies to
\begin{align}
P_{\infty}&(\gamma_{th}, q_{th}) = \nonumber \\
&\sum_{i=1}^{\infty}~\mathbb{P}\left[ C(H_{i}{(t+l,l)}) < \gamma_{th} ~\vert ~ Q(H_{i}{(t,k)}) \leq q_{th}\right]
\nonumber \\ &\times\left[\prod_{i'=1}^{i-1}\mathbb{P}\left[Q(H_{i'}{(t,k)}) > q_{th}\right]\right]\mathbb{P}\left[Q(H_{i}{(t,k)}) \leq q_{th} \right].
\label{eq:infiniteproof4}
\end{align}
Because the resources are identically distributed, \eqref{eq:infiniteproof4} simplifies to
\begin{equation}
P_{\infty}(\gamma_{th},q_{th})  = \sum_{i=1}^{\infty}p~\left( 1 - F_{Q}\left( q_{th}\right)\right)^{i-1}  F_{Q}\left(  q_{th} \right)
\label{eq:first_result}
\end{equation}
where $p = \mathbb{P}\left[ C\left(H_{i}{(t+l,l)}\right) < \gamma_{th} ~\vert ~ Q\left(H_{i}{(t,k)}\right) \leq q_{th}\right]$. The infinite sum in \eqref{eq:first_result} is a geometric series, which reduces to $p$.
This completes the proof for the infinite resource case given by~\eqref{eq:infinite_outage_prob} in Theorem~1.

For the finite resource case, we have~\eqref{eq:finite_proof2} -~\eqref{eq:finite_proof7}, {shown at the top of the page following the next}.
\begin{figure*}[ht]
\begin{align}
P_{\vert \mathcal{R} \vert}\left(\gamma_{th},q_{th}\right)  = & ~\mathbb{P}\left[\bigcup_{i=1}^{\left|\mathcal{R}\right|}\widehat{\mathtt{E}}_{i} \right]=\sum_{i=1}^{\left|\mathcal{R}\right|}\mathbb{P}\left(\widehat{\mathtt{E}}_{i}\right) = \mathbb{P}\left(\widehat{\mathtt{E}}_{\left| \mathcal{R} \right|}\right) + \sum_{i=1}^{\left|\mathcal{R}\right|-1}\mathbb{P}\left(\widehat{\mathtt{E}}_{i}\right) \label{eq:finite_proof2} \\
=&~\mathbb{P}\left[ C\left({H}_{\left|\mathcal{R}\right|}{(t+l,l)}\right) < \gamma_{th}\bigcap_{i'=1}^{\left| \mathcal{R} \right| - 1}Q(H_{i'}{(t,k)}) > q_{th} \right] + \nonumber \\ &  \sum_{i=1}^{\left|\mathcal{R}\right| - 1}\mathbb{P}\left[ C\left(H_{i}{(t+l,l)}\right) < \gamma_{th}\bigcap_{i'=1}^{i - 1}Q(H_{i'}{(t,k)}) > q_{th}\bigcap Q\left(H_{i}{(t,k)}\right) \leq q_{th}\right] \label{eq:finite_proof3} \\
=&~ \mathbb{P}\left[ C\left({H}_{\left|\mathcal{R}\right|}{(t+l,l)}\right) < \gamma_{th}~| \bigcap_{i'=1}^{\left| \mathcal{R} \right| - 1}Q\left(H_{i'}{(t,k)}\right) > q_{th} \right] \mathbb{P} \left[\bigcap_{i'=1}^{\left| \mathcal{R} \right| - 1}Q\left(H_{i'}{(t,k)}\right) > q_{th}\right] + \nonumber\\ &\sum_{i=1}^{\left| \mathcal{R} \right| - 1}\mathbb{P}\left[  C\left(H_{i}{(t+l,l)}\right) < \gamma_{th}~\big\vert \bigcap_{i'=1}^{i - 1}Q\left(H_{i'}{(t,k)}\right) > q_{th}\bigcap Q\left(H_{i}{(t,k)}\right) \leq q_{th} \right] \nonumber  \\
&\times\mathbb{P}\left[\bigcap_{i'=1}^{i - 1}Q\left(H_{i'}{(t,k)}\right) > q_{th}\bigcap Q\left(H_{i}{(t,k)}\right) \leq q_{th}\right] 
\\
=&~\mathbb{P}\left[ C\left({H}_{\left|\mathcal{R}\right|}{(t+l,l)}\right) < \gamma_{th} \right] \prod_{i'=1}^{\left| \mathcal{R} \right| - 1} \mathbb{P} \left[Q\left(H_{i'}{(t,k)}\right) > q_{th}\right] 
\nonumber \\ 
&+ \sum_{i=1}^{\left| \mathcal{R} \right| - 1}\mathbb{P}\left[ C\left(H_{i}{(t+l,l)} \right) < \gamma_{th}~\big\vert~Q\left(H_{i}{(t,k)}\right) \leq q_{th} \right] \nonumber  \\
&\times \left[\prod_{i'=1}^{i - 1} \mathbb{P}\left[Q\left(H_{i'}{(t,k)}\right) > q_{th}\right]\right] \mathbb{P}\left[Q\left(H_{i}{(t,k)}\right) \leq q_{th}\right]
\label{eq:finite_proof4} \\
=& ~\mathbb{P}\left[ C\left({H}_{\left|\mathcal{R}\right|}{(t+l,l)}\right) < \gamma_{th} \right] \left(1 - F_{Q}\left(q_{th}\right)\right)^{\left| \mathcal{R} \right| - 1} + \sum_{i=1}^{\left| \mathcal{R} \right| - 1}p  \left( 1 - F_{Q}\left(q_{th}\right)\right)^{i-1} F_{Q}\left(q_{th}\right)  \label{eq:finite_proof6} \\
=& ~\mathbb{P}\left[ C\left({H}_{\left|\mathcal{R}\right|}{(t+l,l)}\right) < \gamma_{th} \right]  \left(1 - F_{Q}\left(q_{th} \right)\right)^{\left| \mathcal{R} \right| - 1} + p -p  \left(1 - F_{Q}\left(q_{th}\right)\right)  ^{\left| \mathcal{R} \right| - 1}. \label{eq:finite_proof7}
\end{align}
\hrulefill
\end{figure*}
Here, the steps involved in getting to~\eqref{eq:finite_proof4} from~\eqref{eq:finite_proof2} follow similar arguments applied to obtain~\eqref{eq:infiniteproof4} from~\eqref{eq:infiniteproof1}.
Finally, by considering the definition of $p$ above, and recalling that the resources are identically distributed,~\eqref{eq:finite_proof4} can be written as~\eqref{eq:finite_proof6}. Applying the truncated geometric series to~\eqref{eq:finite_proof6}
we obtain~\eqref{eq:finite_proof7}. Now, substituting for the same $p$ from~\eqref{eq:first_result}, we obtain~\eqref{eq:main_outage_expression}. This completes the proof.

\subsubsection*{{$\mathtt{Case~2}$:}}

 {To prove this theorem for $\mathtt{Case~2}$, we utilise the results from $\mathtt{Case~1}$ :
\begin{align}
    P_{\vert \mathcal{R} \vert} &\left(\gamma_{th}, q_{th}\right) \nonumber\\
=&P_1 \left(\gamma_{th}\right) \left(1-F_Q\left( q_{th}\right)\right)^{\left| \mathcal{R} \right| - 1} \nonumber \\ 
&+ P_\infty \left(\gamma_{th}, q_{th}\right)\left(1 - \left(1 -  F_Q\left( q_{th}\right)\right)^{\left| \mathcal{R} \right| - 1} \right).
\end{align}
Clearly, this is expressed in-terms of the critical scenario in which the final resource is chosen, and the non-critical scenario. Because the non-critical scenario for $\mathtt{Case~1}$ and for $\mathtt{Case~2}$ are identical, the second term must be common. Furthermore, it is easy to see from the definition of $\mathtt{Case~2}$ given in~\eqref{eq:rgreedy_case_i} that the probability for the critical scenario of $\mathtt{Case~2}$ is $\mathbb{P}\left[ C(H_{i}{(t+l,l)})\!<\gamma_{th}~|~i = \underset{j\in\mathcal{R}}{\mathrm{argmin}} ~ Q(H_j(t,k)) \right]$, which yields the result.}


\subsection{PROOF OF THEOREM~\ref{thm thm2}  \label{app:A2}}

For Theorem~\ref{thm thm2}, we prove the limiting expressions for $P_{\infty}\left(\gamma_{th}, q_{th}\right)$ given in~\eqref{eq:r_limit_expr}. A similar argument can be applied to \eqref{eq:p_1_limit} and \eqref{eq:q_limit_expr}, which we omit. To determine our limiting expression for $ P_\infty(\gamma_{th}, q_{th})$, with $\mathcal{W}_n$ given by Definition~\ref{def:TPTNFPFN}, consider the sets
\begin{align}
    \mathcal{T}_n = & \left\{ \left( H_{i}\left(t,k\right) , b_{i}\right) \in \mathcal{W}_n~|~Q\left(H_{i}{(t,k)}\right) \leq q_{th},~b_{i} = 1 \right\} \nonumber\\
    \mathcal{V}_n = & \left\{ \left( H_{i}\left(t,k\right) , b_{i}\right) \in \mathcal{W}_n~|~Q\left(H_{i}{(t,k)}\right) \leq q_{th} \right\}. 
\end{align}
Clearly, $\mathcal{T}_n \subseteq \mathcal{V}_n \subseteq \mathcal{W}_n$. Furthermore, $\mathcal{T}_n$ represents the subset of events that will be selected by the predictor \textit{and} result in outage, while $\mathcal{V}_n$ represents the subset of events that will be selected by the predictor and may \textit{or} may not result in outage. By considering \eqref{eq:infinite_outage_prob} of Theorem~\ref{thm thm1}, we have
\begin{equation}
  P_\infty\! \left( \gamma_{th}, q_{th}\right)\! = \mathbb{P}\left[ C\!\left(H_{i}{(t+l,l)}\right)\!<\gamma_{th} |Q\!\left(H_{i}{(t,k)} \right) \!\leq q_{th} \right].\label{eq:probbaility_expressions_for_outage_inf}
\end{equation}
Thus, with \eqref{eq:probbaility_expressions_for_outage_inf} we can estimate the outage probability $P_{\infty}\left( \gamma_{th}, q_{th} \right)$ by taking the ratio of these sets' cardinalities.
Applying the law of large numbers, we have~\eqref{eq:limit_inf_proof0} - \eqref{eq:limit_inf_proof}, shown at the bottom of the {page following the next}. 
\begin{figure*}[!ht]
\begin{align}
   P_\infty \left(\gamma_{th}, q_{th}\right) = & \lim_{n\to\infty}\frac{\left| \mathcal{T}_n\right|}{\left| \mathcal{V}_n\right|} \label{eq:limit_inf_proof0} \\
   = & \lim_{n\to\infty}\frac{{\sum_\mathcal{W}}_n \mathbbm{1} \left( q_{th} - Q\left(H_{i}{(t,k)}\right)\right)b_{i}}{{\sum_\mathcal{W}}_n \mathbbm{1}\left(q_{th} - Q\left(H_{i}{(t,k)}\right)\right)} \label{eq:limit_inf_proof1}\\
   = & \lim_{n\to\infty}\frac{{\sum_\mathcal{W}}_n \mathbbm{1}\left(q_{th}- Q\left(H_{i}{(t,k)}\right)\right)b_{i}}{{\sum_\mathcal{W}}_n \mathbbm{1}\left( q_{th}- Q\left(H_{i}{(t,k)}\right)\right)\left(1 - b_{i}\right) + \mathbbm{1} \left( q_{th} - Q\left(H_{i}{(t,k)}\right)\right) b_{i} }\label{eq:limit_inf_proof2}\\
   = & \lim_{n\to\infty}\frac{\mathrm{FN}_n\left(\mathcal{W}_n, \mathbbm{1}\right)}{\mathrm{TN}_n\left(\mathcal{W}_n, \mathbbm{1}\right) + \mathrm{FN}_n\left(\mathcal{W}_n, \mathbbm{1}\right)},\label{eq:limit_inf_proof}
\end{align}
\hrulefill
\end{figure*}
Here, the final line is obtained from Definition~\ref{def:TPTNFPFN}. This completes the proof for~\eqref{eq:r_limit_expr}. A similar argument can be used to obtain \eqref{eq:p_1_limit} and \eqref{eq:q_limit_expr}, respectively, which we omit here.

\section*{ACKNOWLEDGMENT}
The authors would like to thank Professor Osvaldo Simeone, King's College London, for valuable discussions regarding the system model and main results of this work.

\bibliographystyle{IEEEtran}
\bibliography{IEEEabrv,ref}

\newpage

\begin{IEEEbiography}[{\includegraphics[width=1in,height=1.25 in,clip,keepaspectratio]{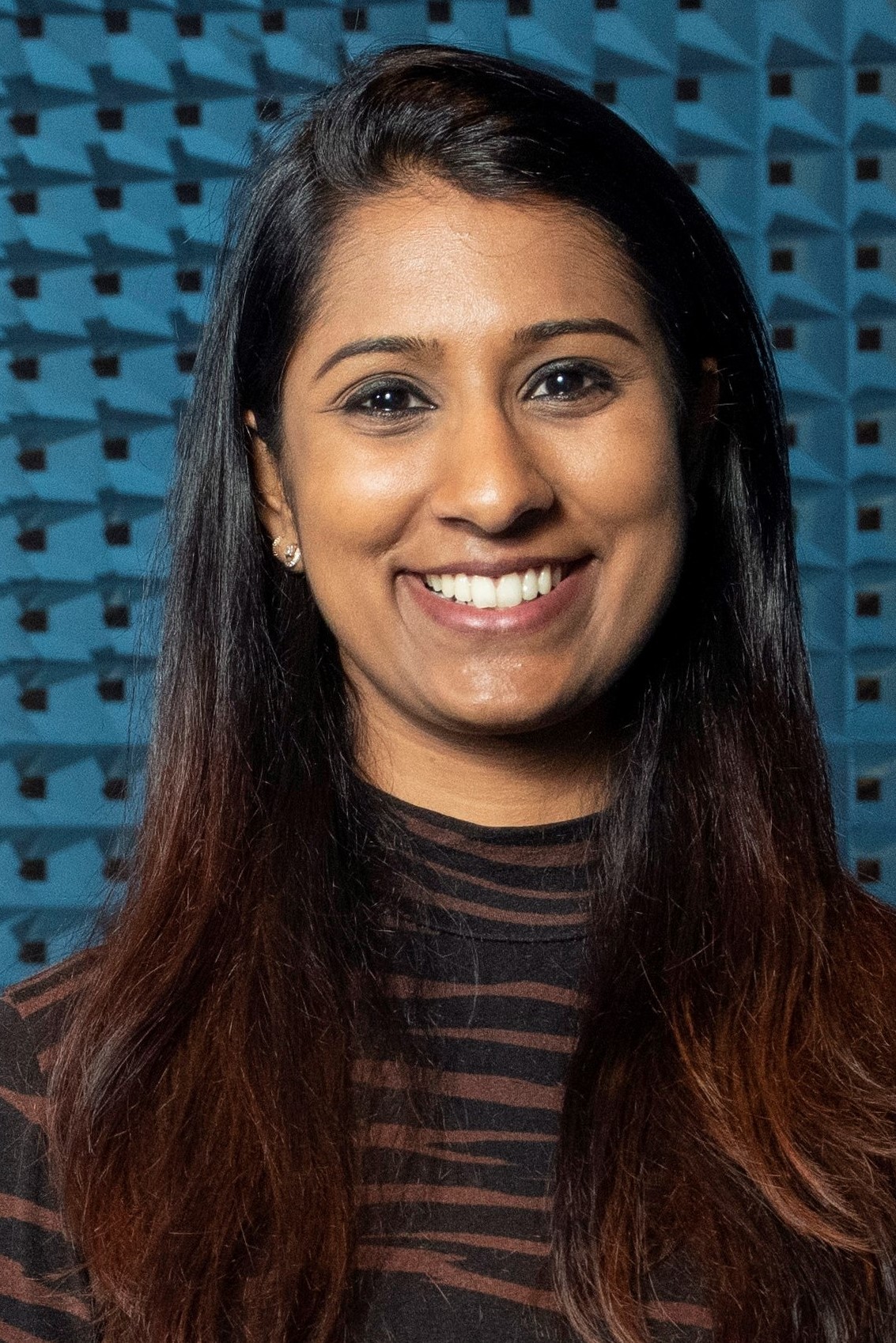}}]{Nidhi Simmons}~(S'14--M'18--SM'23) received the B.E. degree (Distinction) in telecommunications engineering from Visvesvaraya Technological University, Belgaum, India, in 2011, the M.Sc. degree (Distinction) in wireless communications and signal processing from the University of Bristol, U.K., in 2012, and the Ph.D. degree in electrical engineering from Queen's University Belfast, U.K., in 2018. From 2019 - 2020, she worked as a Research Fellow at Queen's University Belfast, U.K. She then received a 5-year research grant worth £500k from the Royal Academy of Engineering, and currently works as a Royal Academy of Engineering Research Fellow at the Centre for Wireless Innovation, Queen's University Belfast. Her research interests include ultra-reliable low-latency communications, channel characterization and modeling, and machine learning for wireless communications. 
\end{IEEEbiography}

\begin{IEEEbiography}[{\includegraphics[width=1in,height=1.25in,clip,keepaspectratio]{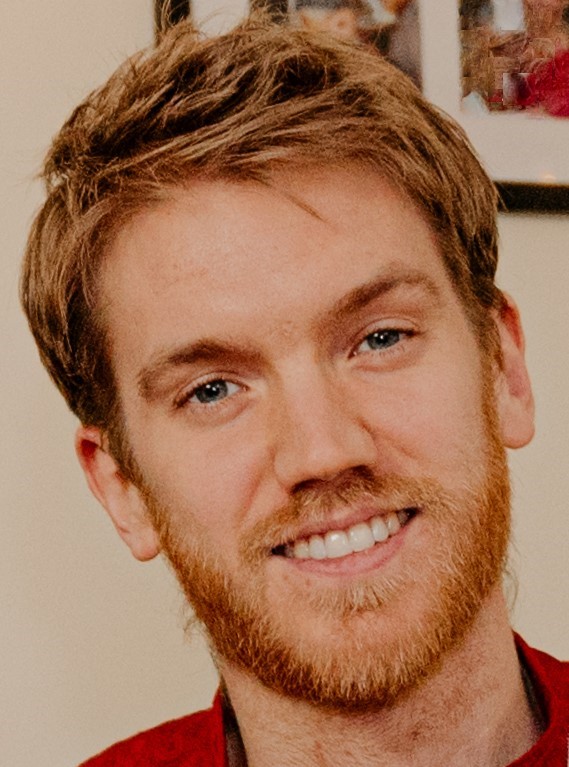}}]{David E. Simmons}~received the BSc in mathematics from the University of Central Lancashire, in 2011, the M.Sc. degree in communications engineering from the University of Bristol, U.K., in 2012, and D.Phil. degree in Engineering from the University of Oxford, U.K., in 2016. His research has focused on studying the information theoretic properties of relay networks as they scale. During his D.Phil. studies, he was a recipient of the Best Paper Award at the 23rd edition of EUCNC’14. From 2016 to 2017, he worked as a PDRA within the Networked Quantum Information Technologies group at the University of Oxford. From 2018 until 2022, he worked as a Senior AI/ML Research Scientist, Senior Software Engineer, and Engineering Team Lead in two startup companies in Belfast, U.K. In January 2023, he co-founded a Web3 technology company, Dhali Holdings Ltd. Since then he has been developing a blockchain enabled open marketplace for on-demand AI. His research interests include communication and network theory, information theory, AI/ML, blockchain and cryptography.
\end{IEEEbiography}

\begin{IEEEbiography}[{\includegraphics[width=1in,height=1.25in,clip,keepaspectratio]{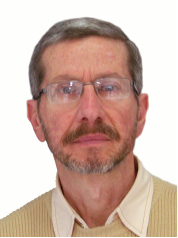}}]{Michel Daoud Yacoub}~was born in Brazil, in 1955. He received the B.S.E.E. and M.Sc. degrees from the School of Electrical and Computer Engineering, State University of Campinas, UNICAMP, Campinas, SP, Brazil, in 1978 and 1983, respectively, and the Ph.D. degree from the University of Essex, U.K., in 1988. From 1978 to 1985, he was a Research Specialist in the development of the Tropico digital exchange family with the Research and Development Center of Telebras, Brazil. In 1989, he joined the School of Electrical and Computer Engineering, UNICAMP, where he is currently a Full Professor. He consults for several operating companies and industries in the wireless communications area. He is the author of Foundations of Mobile Radio Engineering (Boca Raton, FL: CRC, 1993), Wireless Technology: Protocols, Standards, and Techniques (Boca Raton, FL: CRC, 2001), and the coauthor of the Telecommunications: Principles and Trends (Sao Paulo, Brasil: Erica, 1997, in Portuguese). He holds two patents. His general research interest includes
wireless communications. 
\end{IEEEbiography}

\vfill\pagebreak

\end{document}